\documentclass[twocolumn]{revtex4-1}

\usepackage{natbib}
\usepackage{physics} 

\usepackage{enumerate} 
\usepackage{graphicx} %
\usepackage{color}

\usepackage{amsmath, amssymb}

\usepackage[hidelinks]{hyperref}

\usepackage{dcolumn}
\usepackage{bm}
\usepackage{amsmath,amssymb}
\usepackage[squaren]{SIunits}

\usepackage[ruled]{algorithm2e} 
\usepackage{newfloat,algcompatible} 
\usepackage{float}

\hypersetup{
    colorlinks=true,       
    linkcolor=blue,          
    citecolor=blue,        
    urlcolor=blue           
}

\usepackage{soul}

\definecolor{pink}{rgb}{0.858, 0.188, 0.478}
\definecolor{brown}{rgb}{0.59, 0.29, 0.0}
\begin{document}

\preprint{APS/123-QED}


\title{
Magnetic-field-learning using a single electronic spin in diamond \\
with one-photon-readout at room temperature
}

\author{R.~Santagati$^{1,\dagger}$}
\author{A.A.~Gentile$^{1,\dagger}$}
\author{S.~Knauer$^{1,2,\dagger}$}
\author{S.~Schmitt$^3$}
\author{S.~Paesani$^1$}
\author{C.~Granade$^4$}
\author{N.~Wiebe$^4$}
\author{C. Osterkamp$^3$}
\author{L.P.~McGuinness$^3$}
\author{J.~Wang$^1$}
\author{M.G.~Thompson$^1$}
\author{J.G.~Rarity$^1$}
\author{F.~Jelezko$^{3*}$}
\author{A.~Laing$^{1*}$}

\affiliation{$^1$Quantum Engineering Technology Labs, H. H. Wills Physics Laboratory and Department of Electrical and Electronic Engineering, University of Bristol, Bristol BS8 1FD, UK.}
\affiliation{$^2$Centre of Excellence for Quantum Computation and Communication Technology, School of Physics, University of New South Wales, Sydney, NSW 2052,
Australia.}
\affiliation{$^3$Institute of Quantum Optics, Ulm University, 89081 Ulm, Germany.}
\affiliation{$^4$Quantum Architectures and Computation Group, Microsoft Research, Redmond, Washington 98052, USA.}
\affiliation{$^\dagger$These authors contributed equally to this work.}
\affiliation{$^*$email to: fedor.jelezko@uni-ulm.de, anthony.laing@bristol.ac.uk}

\begin{abstract}
Nitrogen-vacancy (NV) centres in diamond
are appealing nano-scale quantum sensors for
temperature, strain, electric fields and, most notably, for magnetic fields.
However, the cryogenic temperatures required
for low-noise single-shot readout
that have enabled the most sensitive NV-magnetometry
reported to date,
are impractical for key applications,
e.g. biological sensing.
Overcoming the noisy readout at room-temperature
has until now
demanded repeated collection of fluorescent photons,
which increases the time-cost of the procedure
thus reducing its sensitivity.
Here we show how machine learning 
can process the noisy readout
of a single NV centre at room-temperature,
requiring on average only one photon per algorithm step, 
to sense magnetic field strength
with a precision comparable to those reported for
cryogenic experiments.
Analysing large data sets from
NV centres in bulk diamond, 
we report absolute sensitivities of 
\unit{60}{\nano \tesla \second^{1/2}}
 including initialisation, readout, and computational overheads.
We show that dephasing times can be simultaneously estimated,
and that time-dependent fields can be dynamically tracked at room temperature.
Our results dramatically increase the practicality of early-term single spin sensors.
\end{abstract}

\maketitle


Quantum sensors are likely to be among the first quantum technologies
to be translated from laboratory set-ups to commercial products~\cite{Taylor:2008cp}.
The single electronic spin of a nitrogen-vacancy (NV) centre in diamond
operates with nano-scale spatial resolution
as a sensor for electric and magnetic fields~\cite{McGuinness:2011ho, Muller:2014fo,Lovchinsky503,Zhao:2011fx,Barry:2016gqa}.
However, achieving high sensitivities for NV-magnetometers has required
a low noise mode of operation available only at cryogenic temperatures,
which constitutes a major obstacle to real-world applications~\cite{Robledo:2011fs,Bonato:2015eu}.
Machine learning has played an enabling role for
new generations of applications in conventional
information processing technologies,
including pattern and speech recognition, diagnostics, and robot control \cite{Murphy:2012uq, Jordan:2015ec}.
Here we show how machine learning algorithms
\cite{Hentschel:2010hm, Granade:2012kj, Wiebe:2014bpbaca, Hincks:2018kg}
can be applied to single-spin magnetometers at room temperature
to give a sensitivity that scales with the Heisenberg limit,
and reduces overheads by requiring only one-photon-readout.
We go on to show that these methods allow multiparameter
estimation to simultanesouly learn the decoherence time,
and implement a routine for the dynamical tracking of time-dependent fields.

Magnetic field sensing with an NV centre
uses Ramsey interferometry
\cite{Taylor:2008cp, Rondin:2014fv, Jelezko:2006jq}.
With a microwave $\pi/2$-pulse the spin vector is rotated
into an equal superposition of its $\sigma_{\textrm{z}}$ spin eigenstates,
such that its magnetic moment
is perpendicular to the magnetic field ($B$)
to be sensed~\cite{Degen:2008jh,Nusran:2012bpa}.
For some Larmor precession time,
$\tau$,
and frequency,
$f_{\textrm{B}} = \gamma B / 2\pi$,
the relative phase between the eigenstates becomes
$\phi = 2\pi f_{\textrm{B}} \tau$,
where $\gamma$ is the electron gyromagnetic ratio 
of magnetic moment to angular momentum.
After a further $\pi/2$-pulse 
to complete the Ramsey sequence,
a measurement of the spin in its $\sigma_{\textrm{z}}$ basis
provides an estimate of $\phi$, the precision of which
is usually improved by repeating the procedure.
Collecting statistics for
a series of different
$\tau$,
produces a fringe of phase varying with time,
from which $B$ can be inferred.

Increasing the sensitivity of a magnetometer
translates to increasing its rate of
sensing precision with sensing time.
The intrinsic resource cost
in estimating $B$ is
the total phase accumulation time~\cite{Arai:2015bsa,
Waldherr:2012kt, Puentes:2014bt},
which is the sum of every $\tau$
performed during an experiment.
A fundamental limitation
on the sensitivity
of an estimate of $B$
is quantum projection noise ---
from the uncertain outcome
of a $\sigma_{\textrm{z}}$-basis measurement ---
the effect of which is conventionally reduced
through repeated measurements,
at the cost of increasing
the total sensing time.
A further typical limitation on sensing precision
is the timescale,
$T^{*}_{2}$,
on which
spin states decohere
due to inhomogeneous broadening
(though spin-echo methods
could extend this
\cite{Balasubramanian:2008ga}).
In an idealised setting,
with an optimal sensing protocol,
the Heisenberg limit (HL)~\cite{Berry:2009fo}
in sensitivity can be achieved,
to arrive at a precision limited by $T^{*}_{2}$
in the shortest time allowed
by quantum mechanics.
In practice, overheads
such as the time required for
initialisation, computation, and readout
must also be accounted for,
while repeated measurements due to experimental inefficiencies 
and low-fidelity readout increase the time to reach
the precision limited by $T^{*}_{2}$.
The increase in total sensing time
due to overheads and repeated measurements 
thus decreases the sensitivity.

A particularly relevant overhead is
the time taken to
readout the state of the spin,
which depends on the experimental conditions.
At cryogenic temperatures,
spin-selective optical transitions
can be accessed such that,
during optical pumping,
fluorescence is almost completely absent
for one of the spin states.
This single-shot method
allows the spin state to be efficiently
determined with a high confidence
for any given Ramsey sequence
(up to collection and detection efficiencies),
resulting in a relatively low readout overhead.
At room temperature, in contrast,
where spin-selective optical transitions
are not resolved in a single shot,
readout is typically performed
by simultaneously exciting a spin-triplet
that includes both basis states,
and observing fluorescence from subsequent decay,
the probabilities for which 
differ by only $\approx35\%$.
Overcoming this classical uncertainty
(in addition to quantum projection noise)
to allow a precise estimate
of the relative spin state probabilities
after a given precession time $\tau$,
required repeated Ramsey sequences to produce
a large ensemble of fluorescent photons.
Such a large readout overhead
significantly reduces the sensitivity
of NV-magnetometry,
and so far, the high sensitivities
reported at cryogenic temperatures
have been out of reach for room temperature operation
by several orders of magnitude.
Yet NV-sensing at cryogenic temperatures
is impractical for biological applications
such as
in-vivo measurements~\cite{McGuinness:2011ho}
and monitoring of metabolic processes~\cite{Degen:2017kw}.

A large body of work
\cite{
Giovannetti:2004jg,Higgins:2007ig,Berry:2009fo,
Higgins:2009fy,Giovannetti:2011jka,Said:2011ex,
Waldherr:2012kt, Nusran:2012bpa,Bonato:2015eu}
has developed and improved quantum sensing algorithms
to surpass the
classical standard measurement sensitivity (SMS).
While the SMS bounds the sensitivity
that can be achieved for NV-magnetometry
with constant phase accumulation time,
phase estimation algorithms using
a set of different precession times $\tau_{\textrm{i}}$,
allow the SMS to be overcome
~\cite{Waldherr:2012kt, Nusran:2012bpa}.
Further improvements in sensitivity are possible
by adapting measurement bases,
to require fewer Ramsey sequences~\cite{Bonato:2015eu}.
However,
sensing algorithms that use a standard Bayesian approach
typically involve
probability distributions
that are
computationally intensive to update,
or which
contain outlying regions that
significantly affect an estimate.
An appealing alternative
\cite{Granade:2012kj, Wiebe:2014bpbaca, Wiebe:2016cn}
uses techniques from machine learning
to approximate a
probability distribution
with a relatively small collection of points,
known as particles.
These methods have been applied to
the problem of learning a Hamiltonian~\cite{Wiebe:2014bpbaca,Wang:2017fre},
and to implement noise-tolerant quantum phase estimation~\cite{Paesani:2017ga}.

Here, we experimentally demonstrate
a magnetic field learning (MFL) algorithm
that operates with
on average only one photon
readout from a single NV centre at room temperature,
and achieves a level of sensitivity
so far only reported
for cryogenic operation~\cite{Bonato:2015eu}.
MFL adapts efficient Bayesian phase estimation
and Hamiltonian Learning techniques
for magnetometry
to achieve a fast convergence
to the correct value of the magnetic field,
and requires no adaptation of measurement bases.
The parameters of our MFL algorithm,
including the number of particles,
can be optimised prior to operation
without adding to the sensing time overhead.
Each precession time $\tau_{\textrm{i}}$
is chosen~\cite{Ruster:2017ee}
as the inverse of the uncertainty $\sigma_{\textrm{i}-1}$
in the algorithm's previous estimate of $B$,
allowing $\tau$ to grow exponentially
to achieve HL scaling in sensitivity.
We tested MFL on a large data set from
$60,000$
Ramsey interferometry experiments
on a bulk diamond NV centre.
We benchmark the performance of MFL against
standard FFT methods, as well as previous experimental results
from other phase estimation algorithms. 
%
Simultaneous to the learning of $B$,
MFL produces an estimate of $T_2^*$,
which, in contrast to other 
phase estimation algorithms,
allows MFL to lower bound its sensitivity
to the SMS, however long its implementation runtime.
Remarkably,
we show that MFL
enables the dynamical tracking
of time-varying magnetic fields
at room temperature.

\begin{figure*}[t]
\centering
\includegraphics[width=\textwidth]{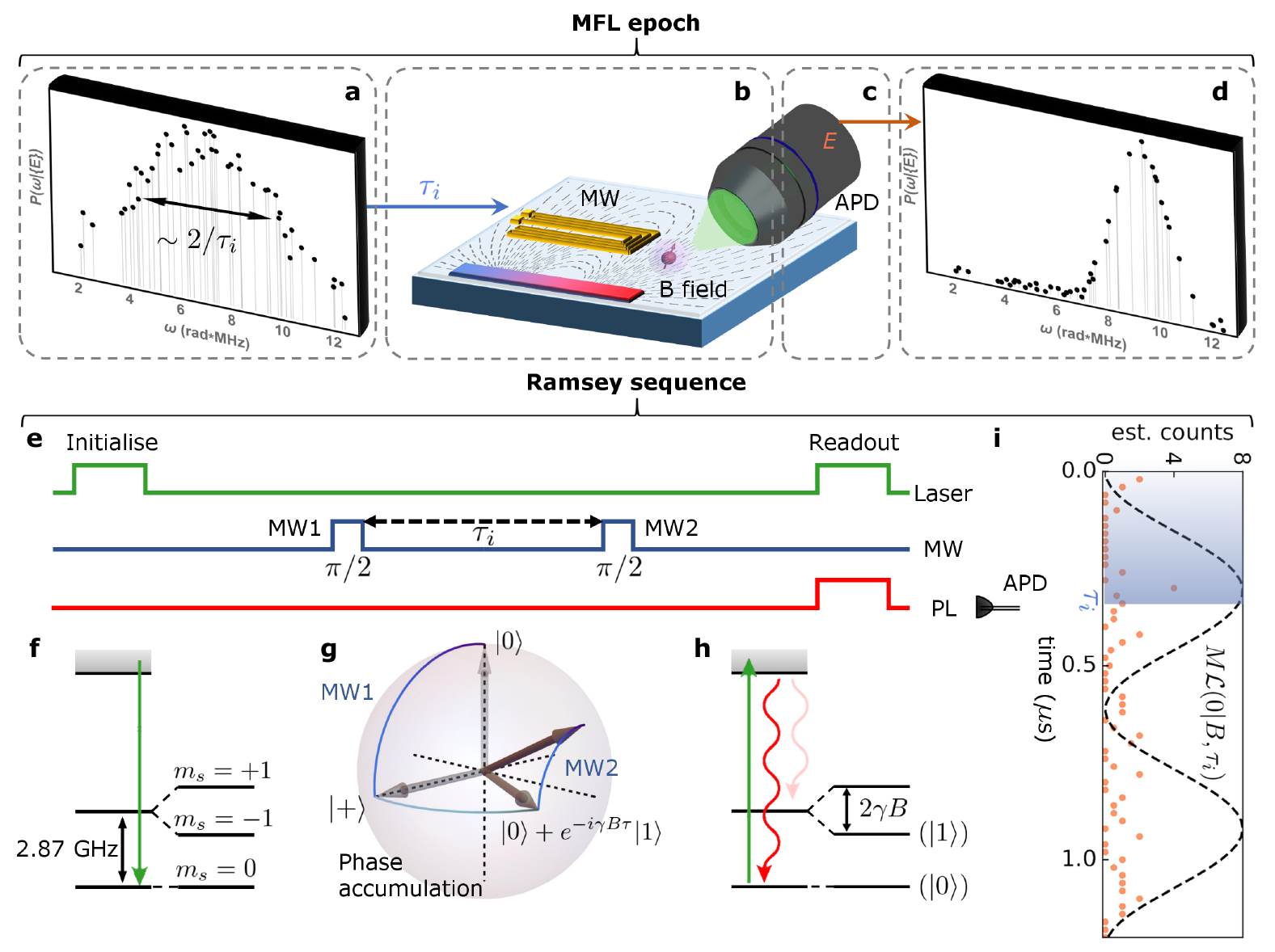}
\caption{
An epoch of the MFL algorithm
including
a Ramsey sequence
and readout.
(a)
The uncertainty encoded in the prior distribution $P_{\textrm{i}-1}$
determines
the phase accumulation time $\tau_{\textrm{i}}$
for the next set of Ramsey sequences.
(b)
A number $M$ of Ramsey sequences are
implemented for $\tau_{\textrm{i}}$,
with the precession
driven by a $B$-field from permanent magnets.
Laser light is focused
with a confocal microscope.
A planar copper wire on the surface of the bulk diamond
generates MW pulses.
(c)
The outcome $E$ from the Ramsey sequences
are measured.
(d)
The prior distribution
is updated $P_{\textrm{i}-1}\rightarrow P_{\textrm{i}}$ through Bayesian inference,
from which the next phase accumulation time, $\tau_{i+1}$ is determined.
(e)
The NV spin vector 
is initialised
with laser light,
rotated with MW pulses,
and, using a second laser pulse,
readout from photoluminescence (PL)
with an avalanche photodiode (APD).
(f)
The electronic energy level triplet
supports initialisation
and MW manipulation
between the
$m_{\textrm{s}}=0$ and $m_{\textrm{s}}=-1$ states,
which encode the basis states,
$|0\rangle$ and $|1\rangle$,
respectively.
(g)
The Bloch sphere depicts the transit of
the electronic state vector for the MW
rotations and Larmor precession.
(h)
Detection
is performed by optically pumping
the basis states to
a higher energy level triplet
and measuring the decay via (non spin-preserving) PL.
(i)
A representative PL fringe
(theory plotted as dashed line)
with orange data-points representing 
the number of detected photons for $M=8$.
}
\label{Fig:Figure1}
\end{figure*}

In general,
Hamiltonian learning algorithms
estimate the parameters $\vec{x}$
of a model Hamiltonian $\hat{H}(\vec{x})$,
through iterations of
informative measurements~\cite{Wiebe:2014bpbaca}.
At each step,
a prior probability distribution $P(\vec{x})$
stores estimates of every parameter and its uncertainty~\cite{Granade:2012kj}.
Similarly,
the four principal recursive steps of MFL,
called an epoch and
depicted in Fig.~\ref{Fig:Figure1}(a-d),
are:
(a)
Choose $\tau_{\textrm{i}}$
for the next Ramsey sequence
from the heuristic
$\tau_{\textrm{i}} \simeq 1/\sigma_{\textrm{i}-1}$,
where $\sigma_{\textrm{i}-1}^2$ is the uncertainty embedded 
in the prior $P(\vec x_{\textrm{i}-1})$.
(b)
Allow the system to evolve under 
$\hat{H}$ for a time $\tau_{\textrm{i}}$,
using the Ramsey sequence shown in
Fig.~\ref{Fig:Figure1}(e-h).
(c)
Measure the outcome $E$,
extracted from
the photo-luminescence
count, e.g. Fig.~\ref{Fig:Figure1}(i).
(d)
Update the prior using Bayes' rule,
$P'(\vec x|E)
\propto \mathcal{L}(E | \vec x; \tau) P(\vec x)$,
where $\mathcal{L}$ is
the likelihood function~\cite{Granade:2012kj}.
%
%
The use of sequential Monte Carlo algorithms
\cite{Granade:2012kj, Wiebe:2014bpbaca, Wiebe:2016cn}
where particles 
are reallocated when required,
makes the inference process practical
and computationally efficient.
Here,
the Hamiltonian for the two relevant NV states
is modelled as
\begin{equation}
\hat{H}(B)= \omega(B)\  \hat{\sigma}_{\textrm{z}} /2= \gamma B \ \hat{\sigma}_{\textrm{z}} /2,
\label{eq:hamiltonian}
\end{equation}
so that $\omega$ is the only parameter
to be estimated to learn the value of $B$.


\begin{figure}[phtb!]
\centering
\includegraphics[
width=0.49\textwidth]{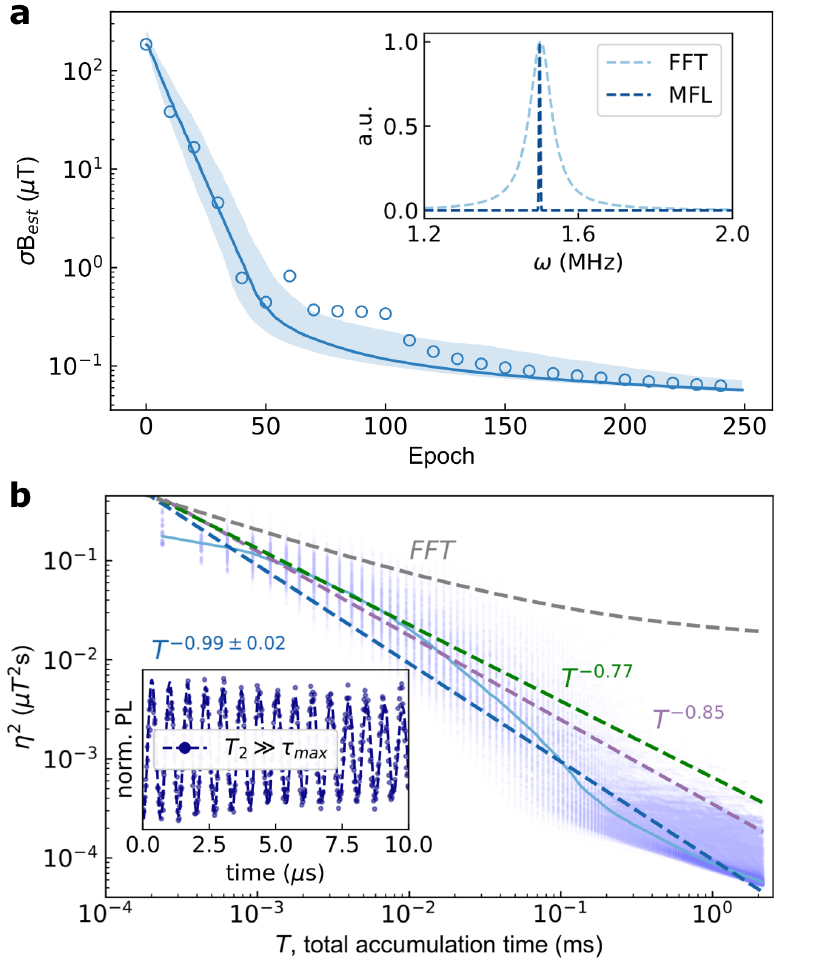}
\caption{
Experimental results for scaling of precision.
Lines represent median values, and performance within the 68.27\% percentile range is shown as shaded areas.
%
(a)
Estimated uncertainty
$\sigma(B_{\textrm{est}})$
is plotted as a function of
the epoch number; data from
one sample run is shown as blue circles.
In the inset, a plot of
the final $\sigma(\omega_{\textrm{est}})$
in the Ramsey frequency
for a typical protocol run,
from FFT (Lorentzian fit)
and MFL (Gaussian fit). 
(b)
The scaling of precision
with
total phase accumulation time $T$,
excluding all overheads,
is shown as density plots
with a linear least-squares fit (blue dashed line).
The FFT approach is plotted as a grey dashed curve.
Scaling for
phase estimation algorithms in Refs.~\cite{Waldherr:2012kt,Nusran:2012bpa}
(respectively green and violet lines) are also reproduced.
The inset shows data from a Ramsey fringe in normalised PL,
with a \unit{20}{ns} sampling rate,
up to $\tau_{\textrm{max}} \sim 0.14 \; T^{*}_{2}$. A least-squares-fit with a decaying sinusoid is shown as a blue dashed line.
%
%
%
}
\label{Fig:Figure2}
\end{figure}

\begin{figure}[phb!]
\centering
\includegraphics[
width=0.5\textwidth]{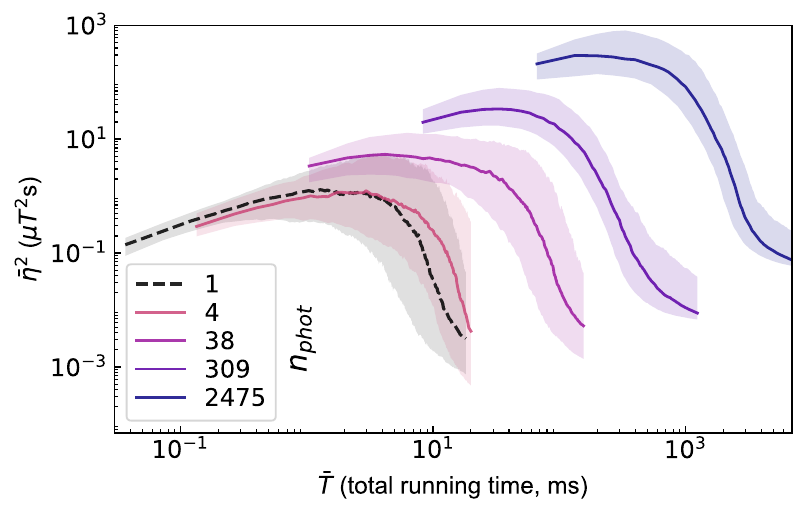}
\caption{
The representative scaling
of precision, including overheads, against total running time is plotted
for different average numbers
of photons detected per epoch (identified by different colours).
Each protocol run for $n_{\textrm{phot}} > 1$ comprises $N=150$ epochs, and only Poissonian noise is modelled in the likelihood function.
For $n_{\textrm{phot}} = 1$, each run comprises $N=500$ epochs and an improved likelihood models also infidelities and losses. 
}
\label{Fig:Figure2C}
\end{figure}

\begin{figure}[t!]
\centering
\includegraphics[width=0.48\textwidth]{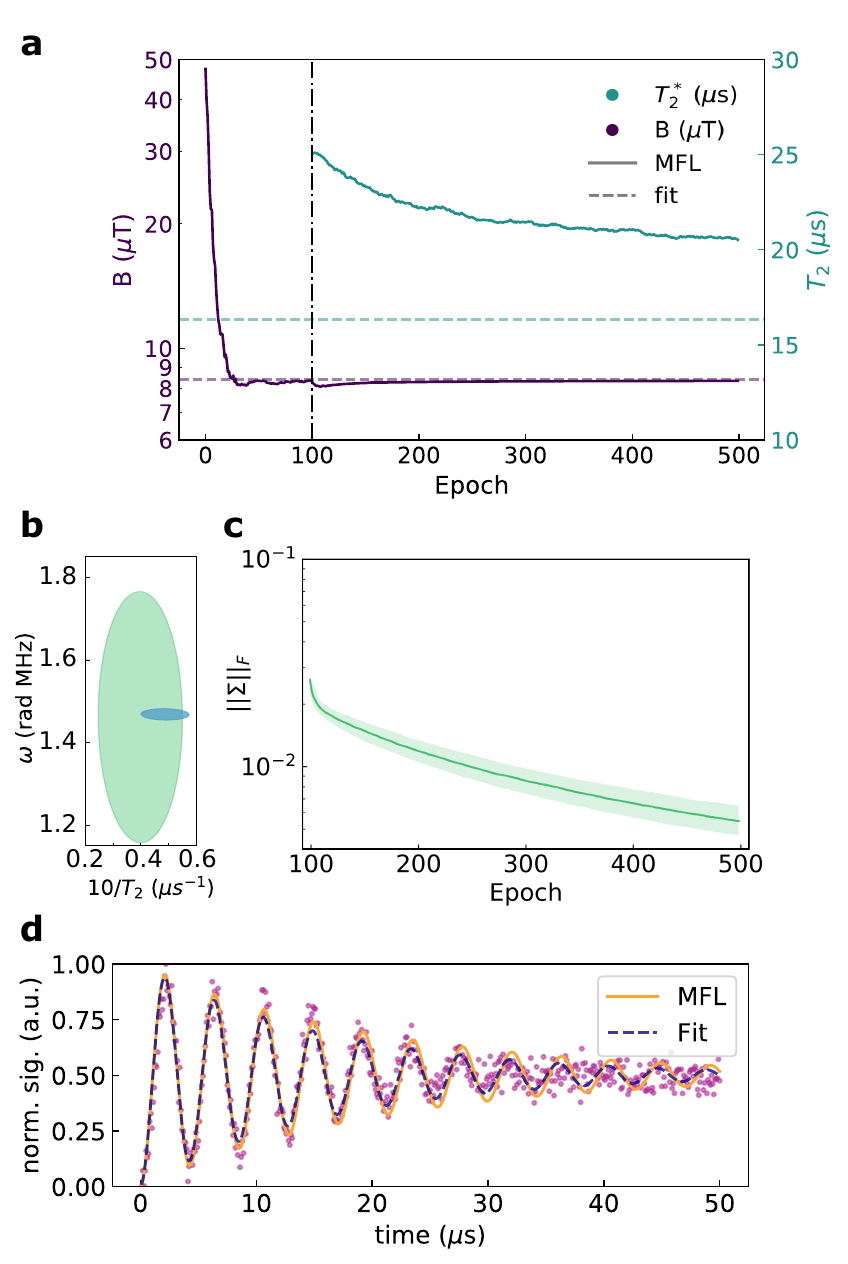}
\caption{
Simultaneous learning of
$T_2^*$ and magnetic field. 
(a)
Simultaneous estimates of
magnetic field $B$ (purple)
and decoherence time $T^*_2$ (green)
for epochs higher than $100$.
Solid lines are from MFL and dashed lines
are from a least squares fit to the Ramsey fringe
data in (d).
(b) 68.27\% credible region
at epoch 100 (green) and 500 (blue)
for $\omega$ and $T^*_2$, reported respectively on the y (x) axes. The smaller area of the distribution at the final epoch indicates the decreased uncertainty on both parameters.
(c)
The norm of the covariance matrix
$\lVert \Sigma \rVert_{\textrm{F}}$,
representing the uncertainty
in simultaneous estimates
of $B$ and $T^*_2$,
is plotted against epoch number.
The median performance is shown as a solid line,
with a shaded area representing the 68.27\% percentile range.
(d)
Renormalised
experimental data for a Ramsey fringe,
along with a least-square-fit
and an MFL-learned decay function
showing decoherence.
}
\label{Fig:Figure3}
\end{figure}

\begin{figure*}[ht!]
\includegraphics[width=0.95\textwidth]{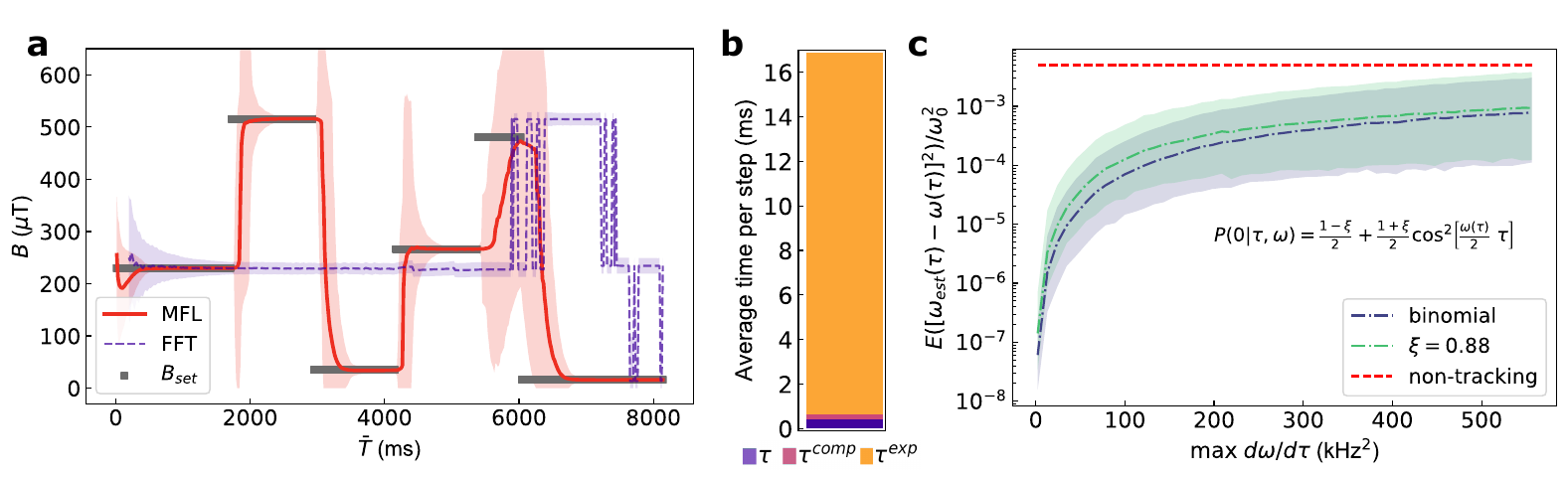}
\caption{
Magnetic field tracking. 
(a) 
Tracking with the MFL protocol
is demonstrated on experimental data,
where step changes in $B$
are indicated by the grey bars
(Here, the number of sequences $M = 4000$).
The solid red line represents typical performances of MFL, with the shaded area indicating performance within a $\sim$68.27\% percentile range.
For comparison, a dashed purple line indicates an FFT protocol applied cumulatively to all data available up to time $\bar T$, with the corresponding uncertainty from a Lorentzian fit as a shaded area of the same colour.
Results after less than 10 data-points are omitted for FFT. 
(b)
Itemisation of the contributions to
the average total time $\bar T$ taken into account in (a):
the precession time $\tau$,
computational ($\tau^{comp}$) and experimental ($\tau^{exp}$) overheads .
(c)
Numerical study of MFL performance in tracking sinusoidally time-dependent magnetic fields $B(\tau)=\omega(\tau)/\gamma$, under ideal conditions ($T_2^* = \infty$, $\tau^{exp} = 0$). 
The y-axis gives the median time-averaged square error 
(nms$_{\omega}$)
in the Ramsey frequency estimate, against the peak speed at which $B$ changes along each simulated Ramsey sequence ($\max d\omega (\tau)/d\tau$). 
The blue dashed line refers to the case including only binomial noise in $\mathcal{L}(B; \tau)$, while the green line is the case with limited readout fidelity ($\xi = 0.88$), as defined in~\cite{Bonato:2017fr}. 
The dashed red line indicates the error 
obtained via a non-tracking strategy.
Shaded areas indicate the $\sim$68.27\% percentile range.
}
\label{Fig:Figure4}
\end{figure*}


Experiments were performed using a confocal set-up,
at room temperature,
with an external magnetic field of
$\approx$\unit{450}{G},
parallel to the NV centre axis,
giving a Zeeman shift of
$\omega = \gamma B$~\cite{Waldherr:2012kt},
where
$\gamma \approx 2 \pi \cross \unit{28}{\mega \hertz /  \milli \tesla}$~\cite{Jakobi:2017fx}.
For each Ramsey sequence,
the electronic spin is
initialised and readout with
\unit{532}{\nano\meter} laser pulses,
by detecting the photoluminescence (PL) signal
with an avalanche photodiode (APD)
for \unit{350}{\nano\second}. The PL signal is then normalised to extract an experimental estimate for $\mathcal{L}$.
%
For every sequence, the experimental overhead is
the sum of the times for
the laser pulses length (\unit{3}{\micro\second}),
an idle time for relaxation (\unit{1}{\micro \second}),
a short TTL pulse for synchronization (\unit{20}{\nano\second})
and the duration of the two MW-pulses
(together $\approx$ \unit{50}{ \nano\second}).

Data for several hundred Ramsey fringes were
generated from experiments on three NV centres,
labelled $\alpha$, $\beta$ and $\epsilon$ (see Supplementary~Table S1).
In particular, the dataset $\epsilon_1$
comprises Ramsey sequences for precession times
increasing from $\tau_{1}$ to $\tau_{500}$
in steps of \unit{20}{ \nano\second}.
For each $\tau_{\textrm{i}}$,
$20275$ sequences were performed,
and data were stored such that
the results from each individual sequence
could be retrieved.
Therefore, $\binom{20,275}{M}^{500}$ subsets of data from $\epsilon_1$
could be selected and combined to construct fringes comprised of
$M$ sequences.
Running MFL on a sample of these subsets allowed its
performance
to be compared over fringes with different
(but fixed within a fringe) numbers of sequences
including down to $M=8$,
where (due to low collection efficiencies)
the average PL count ($n_{\textrm{phot}}$) is approximately one photon.
Additional experiments on the three NVs
generated further data sets for
several hundred fringes that each comprised
tens of thousands of averaged sequences.
%
All implementations of MFL
are reported as
representative behaviour
averaged over $R=$1000
independent protocol runs
(unless otherwise stated)
each using a single fringe
from these data sets.

We begin by analysing how
the estimate of uncertainty
in the magnetic field, $\sigma (B_{\textrm{est}})$,
given by the variance of $P(\vec x)$,
scales with the number of MFL epochs.
For this purpose, we use the dataset $\alpha_1$, with $120$ fringes all obtained with $M=18500$ sequences. 
At every MFL epoch,
given the adaptively chosen
phase accumulation time
$\tau_{\textrm{i}} \simeq 1/\sigma_{\textrm{i}-1}$,
the experimental datum
with $\tau$ minimising ($|\tau -\tau_{\textrm{i}}|$)
is provided to the MFL updater.
Figure~\ref{Fig:Figure2}(a)
shows an exponential decrease in
the scaling of $\sigma (B_{\textrm{est}})$,
until $\sim 50$ epochs are reached.
After this point, the precession times
${\tau}$ selected by MFL saturate
at $\tau_{\textrm{max}}=$ \unit{10}{\micro \second}, and
$\sigma (\textrm{B}_{\textrm{est}})$ is reduced only polynomially fast,
by accumulating statistics
for ${\tau}$ already retrieved.
This slowdown is analogous to
that occurring when the heuristic requires ${\tau}$
exceeding the system dephasing time~\cite{Granade:2012kj}
(see Supplementary Information for details).
A comparison with FFT methods,
inset in Fig.~\ref{Fig:Figure2}(a),
finds that $\sigma(B_{\textrm{est}})$ is
$\sim 40$ times smaller
for MFL.

Neglecting overheads,
the sensitivity $\eta$ of a magnetometer,
is calculated from
\begin{equation}
\eta^2= \delta B^2 = \sigma^2(B_{\textrm{est}}) T,
\label{eq:precision}
\end{equation}
where $T := \sum_i^N \tau_i$ from $N$ epochs.
Figure~\ref{Fig:Figure2}(b)
plots $\eta^2$ against $T$,
for each epoch,
and compares MFL with the
standard FFT method,
using the same $\alpha_1$ set.
The precision of MFL scales as
$T^{-0.99 \pm 0.02}$,
which overlaps with HL scaling
($\propto T^{-1}$).
The FFT method rapidly approaches the SMS
($\propto T^{0}$), 
whereas (neglecting overheads)
the scaling reported for
quantum phase estimation methods
are qualitatively comparable to MFL,
at the expense of more intensive
post-processing~\cite{Said:2011ex}.
%

For a true measure of absolute sensitivity,
experimental and computational overheads
must be accounted for.
Including initialisation, read-out
and computation time, 
into the total running time $\bar T$,
we redefine Eq.~\ref{eq:precision}
for absolute scaling of $\bar \eta$
(see Methods for details).
The average number of luminescent photons, $n_{\textrm{phot}}$,
used for readout during each epoch,
scales linearly with the number of sequences $M$ ($n_{\textrm{phot}}\propto M$);
on average,
one photon every $M\simeq8$ sequences
is detected.
As shown in Fig.~\ref{Fig:Figure2C},
we use MFL to measure
the scaling of $\bar \eta$
with $\bar T$ (up to $250$ epochs)
for decreasing numbers $n_{phot}$
within each epoch.
The plots have a shape characterised by
an initial slow decrease, followed by a
fast increase in precision.
The relatively slow learning rate
for the short phase accumulation times
in the early stages of the algorithm
leads to
a slow increase in phase accumulation time,
since
($\tau_{\textrm{i}} \propto 1/\sigma_{\textrm{i}-1}$).
The algorithm is slowly learning but the total measurement time is increasing faster 
than the decrease in uncertainty.
However,
when the particles start converging to a valid estimate of $B$,
the uncertainty decreases exponentially, overcoming the 
corresponding increase in sensing times.
Our analysis compares well with previous results
performed under cryogenic conditions,
and scaling parameters for linear least squares fitting
obtain a consistent overlap with HL scaling
for protocol update rates up to \unit{13}{Hz}.

Decreasing the number of sequences (thus $n_{\textrm{phot}}$) per epoch
increases the statistical noise,
which extends the slow learning period.
However, the total time $\bar{T}$ decreases
with $n_{\textrm{phot}}$ to produce an increased sensitivity
in a shorter time.
For $n_{\textrm{phot}}<4$, readout infidelities and losses become the dominant noise mechanisms.
%
In the case for $n_{\textrm{phot}}=1$ therefore,
these additional sources of noise were included in the model.
For $n_{\textrm{phot}}=1$ we obtain a sensitivity of
\unit{60}{\nano \tesla \; \second^{1/2}} 
in $\sim$ \unit{10}{ \milli \second} 
(see also the Supplementary Information). 
%

When an NV-sensing algorithm begins to request
precession times $\tau_{\textrm{i}}$ beyond $T^{*}_{2}$,
where no information can be retrieved,
the effectively wasted sensing time
reduces the sensitivity.
Knowledge of $T^{*}_{2}$
can ensure that all
$\tau_{\textrm{i}}$ are less than $T^{*}_{2}$,
to prevent this reduction in sensitivity
and instead guaranteeing it to scale at the SMS
for long sensing times.
Learning
$T^{*}_{2}$ simultaneously with $B$,
as part of a multi-parameter estimation strategy
\cite{Ralph:2017it, Ciampini:2016cf},
can be more efficient
than independently estimating $T^{*}_{2}$ ahead of 
each sensing experiment.
MFL naturally operates as a multi-parameter estimation protocol
when the prior probability distribution $P(\vec{x})$
is multivariate~\cite{Granade:2012kj},
with the uncertainty in its joint probability distribution
captured by a normalised covariance matrix $\Sigma$.

Each precession time $\tau_{\textrm{i}}$ is chosen
proportionality to the inverse of
the (Frobenius) norm of the covariance matrix
(see Methods).
This can incur an initial slow learning period
due to shorter $\tau_{\textrm{i}}$ being initially most useful
in estimating $B$ while longer precession times are
better for an estimation of $T^{*}_{2}$.
We therefore begin MFL in the
single parameter estimation mode for $B$,
and introduce the
simultaneous learning of $T^{*}_{2}$
at epoch $N=100$ (chosen empirically).
%

Figure~\ref{Fig:Figure3} shows results from running
the MFL algorithm on the $\beta_1$ data set,
where $\tau_{max} > T_2^*$.
As is the case for single parameter estimation results,
we find an exponential scaling of
the generalised uncertainty with the number of epochs,
though the learning rate for $B$ is faster
than that for $T^{*}_{2}$.
There is a discrepancy between the estimate of $T^{*}_{2}$
from MFL shown in \ref{Fig:Figure3}(a)
and the fit (non-weighted least-square) to
the decaying sinusoid shown in \ref{Fig:Figure3}(d).
The discrepancy between these two estimates
results from the PGH preferentially requesting $\tau_{\textrm{i}} < T_2^*$,
such that an estimate of $T^{*}_{2}$ is more informed
by data at these relatively shorter time scales
(see Methods).     

The strength of $B$ may not be fixed in time for typical sensing experiments~\cite{Bonato:2017fr}. 
The Bayesian inference process is conceived to learn on-line when experimentally retrieved likelihoods $P(E|\vec x)$
conflict with its prior information. 
Thus, the ability to track time-varying magnetic fields follows naturally from the MFL's processing speed and adaptivity.
With minor controls in the Bayesian inference procedure, MFL can account for such 
fluctuations and high-amplitude changes in the sensed $B$ (See Methods for details). 
%
Here, we test an algorithm that tracks a $B_{\mathrm{set}}$-field
using the $\epsilon_3$ dataset, 
where $B_{\mathrm{set}}$ was experimentally modulated
by changing the position of the permanent magnet
(see Fig \ref{Fig:Figure1}b).
Data recording was paused during magnet adjustments, leading to stepwise transitions in this data set, where the magnetic field instantly jumps to a new strength then remains stable for a period of between hundreds and thousands of milliseconds.

Results are shown in Fig.~\ref{Fig:Figure4}(a),
with a maximum $\approx 30$-fold instantaneous change in $B$.
MFL detects when the posterior distribution
has become non-representative of the most recent measurements,
by increasing the uncertainty, $\sigma (B_{\textrm{est}})$.
After approximately 10 epochs,
the estimate converges to the new value set for $B$.
Figure~\ref{Fig:Figure4}(b) summarises
the different computational and experimental contributions
to the total running time per epoch ($\approx$ \unit{10}{\milli \second}).
The computational time cost of MFL
is $\tau^{comp}\simeq \unit{0.2 }{\milli \second}$,
with the remaining time costs coming from experimental routines.
We note the computational efficiency of MFL
allows a computational overhead ($\tau^{comp}=$\unit{0.21}{\milli\second})
that is smaller than
the average phase accumulation time ($\tau=$\unit{0.41}{\milli\second})
and two orders of magnitude smaller than
the experimental overheads ($\tau^{exp}=$\unit{16.28}{\milli\second}).  

Figure~\ref{Fig:Figure4}(c) shows numerical results
demonstrating the resilience of MFL against
a dynamic component of increasing frequency,
when tracking an A.C. oscillating field
$B(\tau)=\omega(\tau)/\gamma$,
where we choose
$\omega (\tau) = \omega_0 + w \cos(\nu \tau)$,
with $\nu$ a constant and $w \ll \omega_0$.
The effectiveness of the tracking for each run
is captured by a time-dependent normalised squared error
$\textrm{nms}_{\omega} \mathrel
{:=} E [\omega_{\textrm{est}}(\tau)
- \omega(\tau)]^2 / \omega_0^2
= \sum_i^N (\omega_{\textrm{est}}(\tau_i)
- \omega(\tau_i))^2 / (N \omega_0^2)$,
averaged for all $N$ epochs performed,
capturing the efficiency of the tracking as $B$ is not constant along epochs.
Typical estimation errors in $B$ are
lower than 3\% for dynamic components up to \unit{18}{\micro \tesla / \milli \second}. 

The performance of magnetic field learning found for our
room temperature set-up is comparable
to other protocols in cryogenic environments\cite{Bonato:2015eu}.
These methods could be applied to other sensing platforms
where noise has been a limiting factor.
Alternatively, in pursuit of the fundamental limits in absolute sensing precision
they could be used together with
single-shot readout~\cite{Robledo:2011fs}, adaptive measurement bases~\cite{Bonato:2015eu}, faster communication, and dynamical decoupling techniques~\cite{MacQuarrie:2015kg, Farfurnik:2018ip}.
Our methods would be particularly effective in applications where single-spin sensing is desired for nano-scale resolution, but where cryogenic conditions are prohibitive,
such as biological sensing and in new nano-MRI applications~\cite{Barry:2016gqa, Boretti:2016nano}.

\bibliography{sensing}
\clearpage
\newpage

\textbf{Methods}

\begin{footnotesize}

\textbf{MFL execution} 
The data processing was performed by adapting the open \texttt{Python} package \texttt{QInfer} \cite{Granade:2017kb} to the case of experimental metrology.

In order to describe experimental data from Ramsey fringes collected from an NV centre with dephasing time  $T^*_2$, immersed in a magnetic field of intensity $B$, we adopt the likelihood function as in~\cite{Granade:2012kj}:
\begin{equation}
\mathcal{L} (0| B, T_2^*; \tau) = \exp(-\tau/T_2^*) \cos^2 (\gamma B \tau /2) + [1- \exp (-\tau/T_2^*)]/2,
\label{eq:likeT2}
\end{equation}
where $T_2^*$ is a known parameter, or approximated by $T_2^* = \infty$ in all cases when $T_2^* \gg \tau_{max}$.   

In cases when $M>1$, the datum adopted was obtained from $M$ combined sequences as stated in the main text.
Results in Fig.~\ref{Fig:Figure2}a\&b and Fig.~\ref{Fig:Figure4} were all obtained adopting a majority voting scheme to pre-process data from combined sequences~\cite{Paesani:2017ga}. Majority voting decides each single-shot datum according to the most frequent outcome. This is done by previously determining, during the characterisation of the experimental set-up, the average photoluminescence counts ($\bar{n}$) detected throughout the execution of a Ramsey sequence. The datum of a single outcome is determined by comparing the number of photons detected during the measurement (extracted from $M$ sweeps), $n$, and $\bar{n}$. If $n>\bar{n}$ then we set the value of the outcome to $|1\rangle$, otherwise to $|0\rangle$. Without this scheme in place, the outcome of a measurement is assigned sampling from the set $\{  |0\rangle, |1\rangle \}$, with probabilities $P \propto \{ 1-n/n_{max}, n/n_{max} \}$, respectively, with $n_{max}$ the maximum photoluminescence counts estimated during the characterisation.

Other than the study of $\bar \eta$ in Fig.~\ref{Fig:Figure2C}, further examples of the performance of MFL with no majority voting scheme in place are reported in the Supplementary Information.

Errors in the precision scaling are estimated from a bootstrapping procedure, involving a sampling with replacement from the available runs ($R$). The cardinality of each resample matches $R$. The resampling is repeated $\lfloor 0.1 \; R \rfloor $ times. Median precision scalings from each resample are estimated, and the standard deviation from this approximate population of scaling performances is provided as the precision scaling error. 


\textbf{Absolute scaling} 
In Fig.~\ref{Fig:Figure2C} we reported the absolute scaling of $\bar \eta^2 = \sigma^2(B_{\textrm{est}}) \bar{T}$, which requires to take into account the main experimental and computational overheads  contributing to the total running time $\bar T$ of a phase estimation (PE) protocol (communication time $\tau_{comm}$ is not considered here).
In particular, these can be listed as: the time required by the PE algorithm to compute the next experiment $\tau^{comp}$ (here $\sim$ \unit{0.4 }{\micro \second} per step, per particle on a single-core machine), the duration of the laser pulse $\tau^{las}$ for initialisation and readout (\unit{3}{\micro \second} in total), the waiting time $\tau^{wait}$ for relaxation (\unit{1}{\micro \second}), a short TTL pulse $\tau^{TTL}$ for the photodetector (\unit{20}{\nano \second}) and the duration $\tau^{MW}$ of MW-pulses (approximately \unit{50}{\nano \second} in total). 
Including variable and constant overheads, we obtain:
\begin{equation}
\bar{T} = \sum_i^N ( \tau_i + \tau^{comp}_i) + N M (\tau^{las} + \tau^{wait} + \tau^{TTL} + \tau^{MW})
\label{eq:tottime}
\end{equation}
after $N$ epochs of a PE algorithm. 


In the $n_{\textrm{phot}} = 1$ case, the final $\sigma(B_{\textrm{est}}) \simeq$ \unit{0.45}{\micro \tesla} after $500$ epochs, and $\bar{T} \simeq $ \unit{18}{\milli \second}, that is $\bar \eta \simeq$ \unit{60}{\nano \tesla \second^{1/2}}.
In the $n_{\textrm{phot}} = 20$ case, exhibiting a precision scaling that is essentially Heisenberg limited, the uncertainty saturates at protocol convergence ($\sim 150$ epochs) to $\sigma(B_{\textrm{est}}) \simeq$ \unit{0.3}{\micro \tesla}, for a total running time $\bar{T} \simeq $ \unit{78}{\milli \second}. This leads to a final sensitivity $\bar \eta \simeq$ \unit{84}{\nano \tesla \second^{1/2}} and \unit{12.8}{\hertz} repetition rates.



\textbf{Multi-parameter Learning} 
For the multi-parameter case, we use again Eq.~\ref{eq:likeT2}, but now considering the unknown parameters $\vec{x} = \{B, T_2^*\}$.
Each precession time $\tau_{\textrm{i}}$ is chosen
proportionally to the inverse of
the Frobenius norm of the covariance matrix,
$\Vert \Sigma \Vert_F = \Vert cov(B/b, T_2^*/t_2) \Vert_F$.
The parameters $b = \max_{B: P(B) \neq 0} B $ and $t_2 = \min_{T_2^*: P(T_2^*) \neq 0} T_2^* $ are introduced to render $\Vert \Sigma \Vert_F$ dimensionless, with $P$ the prior at epoch $N=100$, when both parameters start to be learnt simultaneously. In this analysis this corresponded to $b = $ \unit{11}{\micro \tesla} and $t_2 = $ \unit{20.2}{\micro \second}, however we stress how different choices would be possible, with equivalent results for $\Vert \Sigma \Vert_F$, up to a  normalisation factor.
We observed that MFL estimates of the dephasing time may differ consistently from a non-weighted least-square fit. In the presence of dephasing, the heuristic of MFL will preferentially adopt experiments with $\tau < T_2^*$. This relation is similar to a weighing mechanism of the data (see also Supplementary~Information
), preferring more consistent observations. 
On the other hand, a least-square fit will attempt to equally mediate over data-points where the contrast in the fringes is almost completely lost, underestimating $T_2^*$.


\textbf{MFL tracking} 
We mentioned that Bayesian inference processes are ideally suited for tracking purposes.
However, we observe that in cases where the magnitude of the changes in the parameter $\vec{x}$ completely invalidates the a-posteriori credibility region, the recovery time of a standard Hamiltonian learning protocol might be unsuitable for practical applications. 
To tackle also this situation, we modified here the standard update procedure to reset its prior when the effective sample size of the particles' ensemble is not restored above the resampling threshold by a sequence of resampling events. Details and a pseudocode are provided in Supplementary~Information.


\textbf{FFT execution} 
For most analyses, FFT estimates were run against the whole datasets available. 
For example in the case of Fig.~\ref{Fig:Figure2}, the final estimate provided by a single run of FFT was performed using once all of the 500 phase accumulation times, recorded with \unit{20}{\nano \second} steps, for a representative subset among those available in $\alpha_1$ (Supplementary~Table~1). 
We emphasise how this amounts to twice as many $\tau$'s as those actually used by the MFL algorithm (being the single-run estimate reported as converged after 250 epochs).

The only exception is the tracking in Fig.~\ref{Fig:Figure4}, where the data-points were cumulatively added to the dataset. 
In such tracking applications, as long as $B$ is kept constant, the estimate from FFT compares to MFL in a way similar to Fig.~\ref{Fig:Figure2}.
However, FFT keeps estimating $B$ from the prominent peak in the spectrum, corresponding to the $\omega$ that was maintained for the longest time, not the most recent. Thus, it fails to track changes as they occur. 


\textbf{Experimental details} 
In Ramsey interferometry, as performed here, we measure the magnetic field component parallel to the NV centres' symmetry axes. However, the MFL protocol can be expanded to differently orientated NV centres, to detect arbitrary orientated magnetic fields.

The experiments are performed here using two different $^{12}$C isotopically purified diamond samples. For the Ramsey interferometry we use the $\text{m}_\text{s}=0$ and $\text{m}_\text{s}=-1$ electronic sublevels.


\textbf{Photon number estimation}
After exciting a single NV centre by a 532nm laser pulse, the red-shifted, individual photons were detected by an avalanche photodiode. A time-tagged single photon counting card with nanosecond resolution was used for recoding. A TTL-connection between the time-tagger and the MW pulse generator synchronises the photon arrival time with respect to the pulse sequence and allows to record the number of detected photons for every single laser pulse.
Thereby, the photon detection efficiency is mainly limited by the collection volume, the total reflection within diamond (due to the high refractive index) and further losses due to the optics. This results in a photon detected about every eighth laser pulse. Thus, to readout the NV state with high-fidelity (and about 30\% contrast) multiple measurements are usually required for meaningful statistics.

\textbf{Acknowledgements}
The authors thank Cristian Bonato and Marco Leonetti for useful discussion.
M.G.T. acknowledges support from the ERC starting grant ERC-2014-STG 640079 and and an EPSRC Early Career Fellowship EP/K033085/1. 
J.G.R. is sponsored under EPSRC grant EP/M024458/1. 
L.M. and F.J. would like to acknowledge DFG (Deutsche Forschungsgemeinschaft grants
SFB/TR21 and FOR1493), BMBF, VW Stiftung, European
Research Council DIADEMS (Simulators and Interfaces with Quantum
Systems), ITN ZULF and BioQ.
A.Laing acknowledges support from an EPSRC Early Career Fellowship EP/N003470/1.

\end{footnotesize}

\clearpage
\newpage

\appendix

\newcommand{\hbAppendixPrefix}{S}
\renewcommand{\thefigure}{\hbAppendixPrefix\arabic{figure}}
\setcounter{figure}{0} 

\renewcommand{\thetable}{\hbAppendixPrefix\arabic{table}} 
\setcounter{table}{0}
\renewcommand{\theequation}{\hbAppendixPrefix\arabic{equation}} 
\setcounter{equation}{0}

\onecolumngrid
\begin{center}
\large{\textbf{
Supplementary Information}}
\end{center}

\section*{Classical and quantum likelihood estimation}
\label{sec:QCLE}

In the main text we have introduced the Bayesian inference process underlying the MFL protocol (known as CLE, Classical Likelihood Estimation\cite{Granade:2012kj}) as composed by four main steps.
Here we expand the discussion to provide additional details and comments about the adoption of CLE.


\begin{enumerate}

\item At each epoch the prior distribution $P(\vec x)$ is used to choose what experimental setting to use for the next iteration. In MFL, the only experimental setting is the phase accumulation time $\tau$, that can be updated effectively using the so called particle guess heuristic (see the section below). 
\item The quantum system undergoes an appropriate evolution, according to the Hamiltonian $\hat H$.  
	\begin{enumerate}
	\item The system is prepared in an appropriate initial state $\ket{\psi}$, chosen such to have informative evolution under $\hat H$. E.g. a state orthogonal to the Hilbert subspace spanned by the Hamiltonian eigenstates. We remark how $\ket{\psi}$ is not adaptive in CLE. In this work, the NV centre is always prepared in $\ket{\psi} = \ket{+}$.
	\item Let the system evolve according to its Hamiltonian $\hat{H}$ for the chosen time $\tau$.
	\end{enumerate}
\item  A measurement is performed on the system (here the quantum sensor).  In MFL we perform a projective measurement on the $\{ \ket{0}, \ket{1} \}$ computational basis, obtaining a bipartite outcome $E \in \{0,1\}$. 
	\item The computed likelihoods are used to update the probability distribution of the Hamiltonian $\vec{x}$ parameters
	\begin{enumerate}
	\item The same experiment is also performed on a simulator, implementing a generic parametrised Hamiltonian $\hat{H}(\vec{x})$, thus providing an estimate for the likelihood $P(E|\vec{x},\tau)$, i.e. the probability of measuring outcome $E$ when $\vec x$ is chosen as parameter.
    \item It is thus possible to apply Bayes' rule:
    \begin{equation}
    P'(\vec{x}|E)=\frac{P(E|\vec{x},\tau)P(\vec{x})}{P(E)},
    \label{eq:SI_bayesupdate}
    \end{equation}
    where $P(\vec{x})$ can be immediately inferred from the prior at the corresponding epoch, while $P(E)$ is a normalization factor.
	\end{enumerate}

\end{enumerate}

Steps 1 -- 4 are repeated until the variance of the probability distribution $\sigma(\vec{x}_{\textrm{est}})$ converges, or falls below a pre-definite target threshold.
In cases with limited readout fidelity like for the NV centre set-up presented here, in step 3 a meaningful statistics might be cumulated for $E$ repeating the same measurement a number of times $M$, as suggested in the main text. Evidences from the text suggest that in most cases, this is a suboptimal choice for the absolute scaling performance of the MFL protocol. 
Note how only steps 2 \& 3 involve the quantum sensor. All other steps require instead a simulator. In particular, step 4(a) can be performed on a classical or quantum simulator, the choice of the second being justified whenever the size of the sensor, and the eventual lack of an analytical model to simplify the evolution, make the system simulation classically expensive. In this case, the inference process is known as Quantum Likelihood Estimation (QLE, \cite{Wiebe:2016cn}).


\begin{table}[ht!]
    \centering
\begin{tabular}{|c|c|c|c|c|c|c|c|}
	\hline 
    label& NV centre & 
    \begin{tabular}{c} 
    sets (n) 
    \end{tabular} &
    \begin{tabular}{c}
    sequences (M) 
    \end{tabular} & 
    $\Delta \tau$ (ns) &
    $\bar{B}_{\textrm{est}}$ ($\mu$T) & 
    \begin{tabular}{c}
    $T^{*}_{2}$ \\ (from fit, \unit{}{\micro \second})
    \end{tabular} & 
    $\eta$ ($\bar{\eta}$) \\ 
    \hline
	\rule{0pt}{2.5ex}$\alpha_1$ & $\alpha$ & 120 & 18500 & 20 & 52  & - & $T^{-0.99 \pm 0.02}$ (-) \\ 
	$\alpha_2$ &      & 115 & 18500 & 20 & 710  & - & $T^{-0.97 \pm 0.03}$ (-)  \\
    
    \hline 
 \rule{0pt}{2.5ex}$\beta_1$ & $\beta$  & 67 & 30000 & 100 & 8.3 & 16 & - (-)  \\ 
    
     \hline 
    
 \rule{0pt}{2.5ex}$\epsilon_1$ & $\epsilon$    & 1 & 20275  & 20 & 58 & - & - (
            $T^{-0.73 \pm 0.03} \rightarrow T^{-1.0 \pm 0.02}$) \\
  	$\epsilon_2$ &      & 1 & 8876  & 200 & 6.0 & 64 & $T^{-0.91 \pm 0.03}$ (-)  \\
    $\epsilon_3$ &     & 1 & 44000 & 20 & $10 \rightarrow 550$ & - & - (-) \\ 
 	
    \hline
    
\end{tabular} 
\caption{
\textbf{Synopsis of available data}. 
Table summarising the different data sets and systems used in this analysis, along with representative MFL performances in precision scaling. Datasets $\alpha_{2}, \epsilon_{2}$ are discussed only in the Supplementary~Information. 
}
\label{fig:SI_table}
\end{table}

\begin{figure*}[ht!]
\centering
\includegraphics[width=0.85\textwidth]{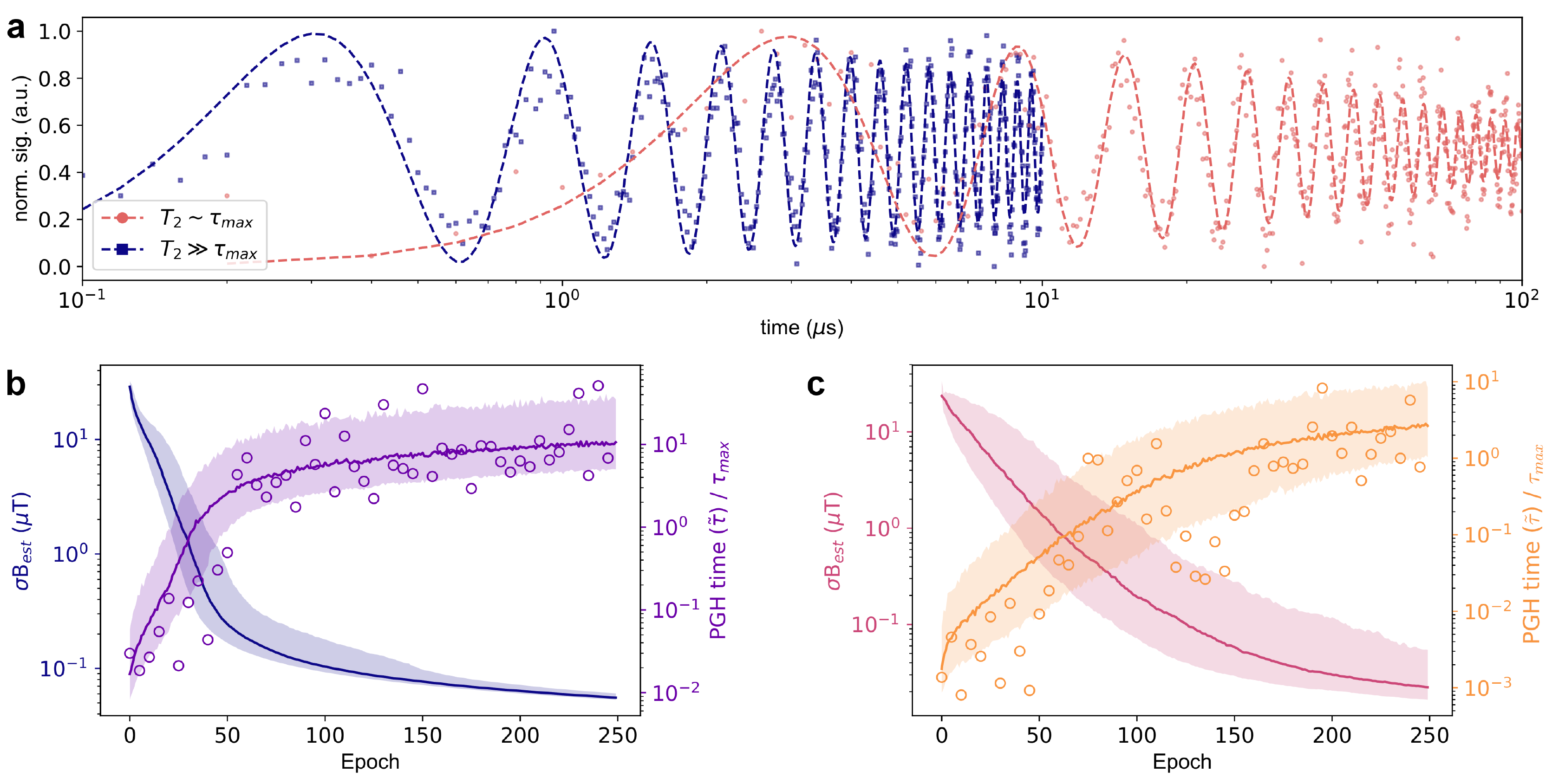} 
\caption{
\textbf{Analysis of the behaviour of the PGH} for datasets where $T^{*}_{2} \gg \tau_{max}$ or $T^{*}_{2} \simeq \tau_{max}$, reported in the plots in darker and brighter colours, respectively. The first dataset is collected with $B \simeq 58$ \unit{\micro \tesla}, whereas the second has $B \simeq 6$ \unit{\micro \tesla}.  
\textbf{a}, Renormalized photon counts along two different Ramsey experiments with the same NV centre (scatter plots). Superimposed a least-square fit (dashed lines), adopting the oscillatory function with depolarizing noise as in Eq.~\ref{eq:likeT2} of Methods. 
\textbf{b}, Estimated uncertainty $\sigma (B_{\textrm{est}})$ and ratio between PGH-generated time and $\tau_{max}$ as available from the first dataset, plotted against each epoch of the MFL algorithm. A majority voting method is adopted, under the hypothesis that $T^{*}_{2} \gg \tau_{max}$. Solid lines are median values calculated over 1000 independent runs, whereas shaded areas are 68.27\% percentile ranges centred around the median. Superimposed as a scatter plot, a sample of times generated by the PGH during an representative run.  
\textbf{c}, Same as in \textbf{b}, for the case where $T^{*}_{2} \simeq \tau_{max}$. No majority voting is in place, and data from the experiment are extracted probabilistically from the experimentally estimated likelihoods.
}
\label{fig:SI_timesaturatePGH}
\end{figure*}


\section*{Sequential Monte Carlo approximation and Particle Guess Heuristic}
\label{sec:SI_smcpgh}

The protocol performances in terms of computational overhead are made possible by adopting advanced approximate Bayesian inference methods. In particular, MFL inherits from CLE the Sequential Monte Carlo (SMC) approximation \cite{Granade:2012kj,Wiebe:2014bpbaca,Hincks:2018kg}.
Within this approximation only a finite number of values $\vec{x}_i$ (called particles) are sampled from the prior distribution, and thus used to approximate the prior in each update step. This approximation makes the Bayesian update as in Eq.~\ref{eq:SI_bayesupdate} numerically feasible. 

If the particle positions $\vec x_i$ were held constant throughout the inference process, and starting the protocol from a uniform prior, the cardinality of their set should scale approximately as $n_{\textrm{part}} := | \{ \vec x_i \} | \propto \Delta B / \tilde \sigma(B_{\textrm{est}})$, where $\Delta B$ is the expected magnetic field range to be sensed, and $\tilde \sigma(B_{\textrm{est}})$ is the targeted uncertainty upon convergence. 
This is inefficient, as with the inference progressing through epochs, many particles will provide very limited support to the updated prior approximation. Indeed, for a successful learning process $\lim_{N \rightarrow \infty} P(\vec{x}_i) \rightarrow 0$ for the weights $w_i \propto P(\vec{x}_i)$ of most particles, as they have been effectively ruled out by the observations. 

This inefficiency can be addressed with resampling methods, that allow the particles to be sampled again from the updated posterior, whenever their weights signify that the effective size of the sampled particles $\sum_i w_i^2$ has fallen below a user-defined threshold. 
Following \cite{Granade:2017kb}, here we adopt a Liu-West resampler (LWR) with optimised resampling threshold $\tt{t}$$_{resample} = 0.5$ and smoothing parameter $\tt{a}$$=0.9$. These parameters allow to tune when and to what extent the positions of the particles can be altered by the LWR~\cite{Granade:2012kj}. Hence, it was possible to accurately represent $P(\vec x)$ throughout the whole protocol execution, whilst employing not more than $1500$ particles for the discretisation in most cases. The only exceptions were limited fidelity cases, as for the absolute scaling we chose the number of particles $n_{part}$ according to the empirical rule
\begin{align}
&n_{part} = 25000/(\log(M)+1) \nonumber \\
&{\tt{t}}_{\textrm{resample}} = \max(0.1, 0.5-0.4/\log(M) ) \nonumber \\
&{\tt{a}} = 0.9+0.08 \; M/M^{max},
\label{eq:SI_numparticles}
\end{align}
with $M$ the number of averaged single sequences selected. This increase in the number of particles can be justified by a corresponding reduction in the risk of ``aggressive'' resampling leading to inference failures, heralded especially by multi-modality in the parameter distribution~\cite{Hincks:2018kg}.

The particle guess heuristic (PGH) plays a fundamental role in the effectiveness of the MFL protocol. PGH was introduced in \cite{Wiebe:2014bpbaca} to provide optimal choice of the experimental $\tau$ (here the phase accumulation time) in analytically tractable cases of Hamiltonian Learning protocols. Such cases happen to include the sensing Hamiltonian presented in the main paper as Eq.~\ref{eq:hamiltonian}. The PGH samples two particles $\{\vec{x}_0, \vec{x}_1\}$ from the particle distribution $P(\vec{x}_i)$, and then chooses:
\begin{equation}
 \tau = \frac{1}{\Vert \hat{H}(\vec{x}_0) - \hat{H}(\vec{x}_1) \Vert} .
\end{equation}
In the single parameter where only $B$ is sensed, $\tau \simeq 1/ \sigma(B_{\textrm{est}})$, where $\sigma(B_{\textrm{est}})$ represents the standard deviation of the Gaussian-approximated posterior distribution $P(B)$. Intuitively, this corresponds to selecting longer, more informative accumulation times, as the estimated uncertainty about the parameter to be learned shrinks.

\section*{The role of $\mathbf{T_2^*}$ in time adaptivity}
\label{sec:SI_T2}

In the main text, we observed the emergence of a slowdown in the learning rate, when MFL chooses accumulation times $\tau_i \geq \tau_{max}$. This slowdown appears when plotting either the scaling in $\sigma(B_{\textrm{est}})$ as well as $\eta$ (respectively Fig.~\ref{Fig:Figure2}a\&b, referring for example to the dataset $\alpha_1$). 
This dataset represents a situation, where $\tau_{max} \ll T^{*}_{2}$ is chosen as a maximum time budget per-epoch. In this case, once the PGH encounters the $\tau_{max}$ limit, learning by statistical accumulation of data-points with $\tau \simeq \tau_{max}$ will occur, and MFL precision scaling will tend to $\eta^2 \propto 1/\sqrt{\tau_{max} T}$. We highlight this behaviour, using averaged sequences from the whole set $\epsilon_1$, in Fig.~\ref{fig:SI_timesaturatePGH}b, plotting the scaling in $\sigma(B_{\textrm{est}})$ alongside with the ratio $\tau/\tau_{max}$. When the plateau in  $\sigma(B_{\textrm{est}})$ occurs, we correspondingly observe that a typical run of MFL starts suggesting $\tau \geq \tau_{max}$, before saturating as the uncertainty converges. 

Note, this artefact deriving from the artificial choice of a maximum time budget $\tau_{max}$ is equivalent to the phenomenon exhibited in correspondence of dephasing noise, reducing the contrast from experimental Ramsey fringes, like in the data reported in Fig.~\ref{Fig:Figure3}. To prove it, we show in Fig.~\ref{fig:SI_timesaturatePGH}c the same performance for the dataset $\epsilon_2$, where $\tau_{max} =$ \unit{100}{\micro \second} $\simeq 1.6 \; T^{*}_{2}$ (estimated from a least-squares fit), and $B=$\unit{6}{\micro \tesla} to have approximately the same number of periods in the corresponding Ramsey fringe, as in dataset $\alpha_1$ (refer to Fig.~\ref{Fig:Figure3}). For MFL to deal properly with decaying data, in this analysis we remove any majority voting scheme from the data processing, and at each epoch the corresponding datum $E$ is probabilistically extracted from the experimentally estimated likelihood (see Methods). This justifies the slowdown in the scaling of $\sigma(B_{\textrm{est}})$, as each data-point is now affected by the same amount of binomial noise that would occur in a set-up with the same readout fidelity, but single-shot measurements. In other words, the additional information acquired about the likelihood $\mathcal{L} (B; \tau)$ by combining $M \gg 1$ sequences for each measurement is partially removed from the inference process by the bipartite $E$ sampling. For this case, we observe the plateau in $\sigma(B_{\textrm{est}})$ occurs when the adaptive choice of the phase accumulation time saturates in average to $\tau \simeq T_2^*$ (though a single run will oscillate around this value, as emphasised by the behaviour for a single run also reported in Fig.~\ref{fig:SI_timesaturatePGH}c).
Similarly, also the scaling in precision plateaus when $\tau \simeq T^{*}_{2}$ (not reported for brevity), slowing towards $\eta^2 \propto 1/{T^{*}_{2}}$. A formal discussion of this saturation is performed in the following section.



 \begin{figure*}[hbt!]
\centering
\includegraphics[width=0.85\textwidth]{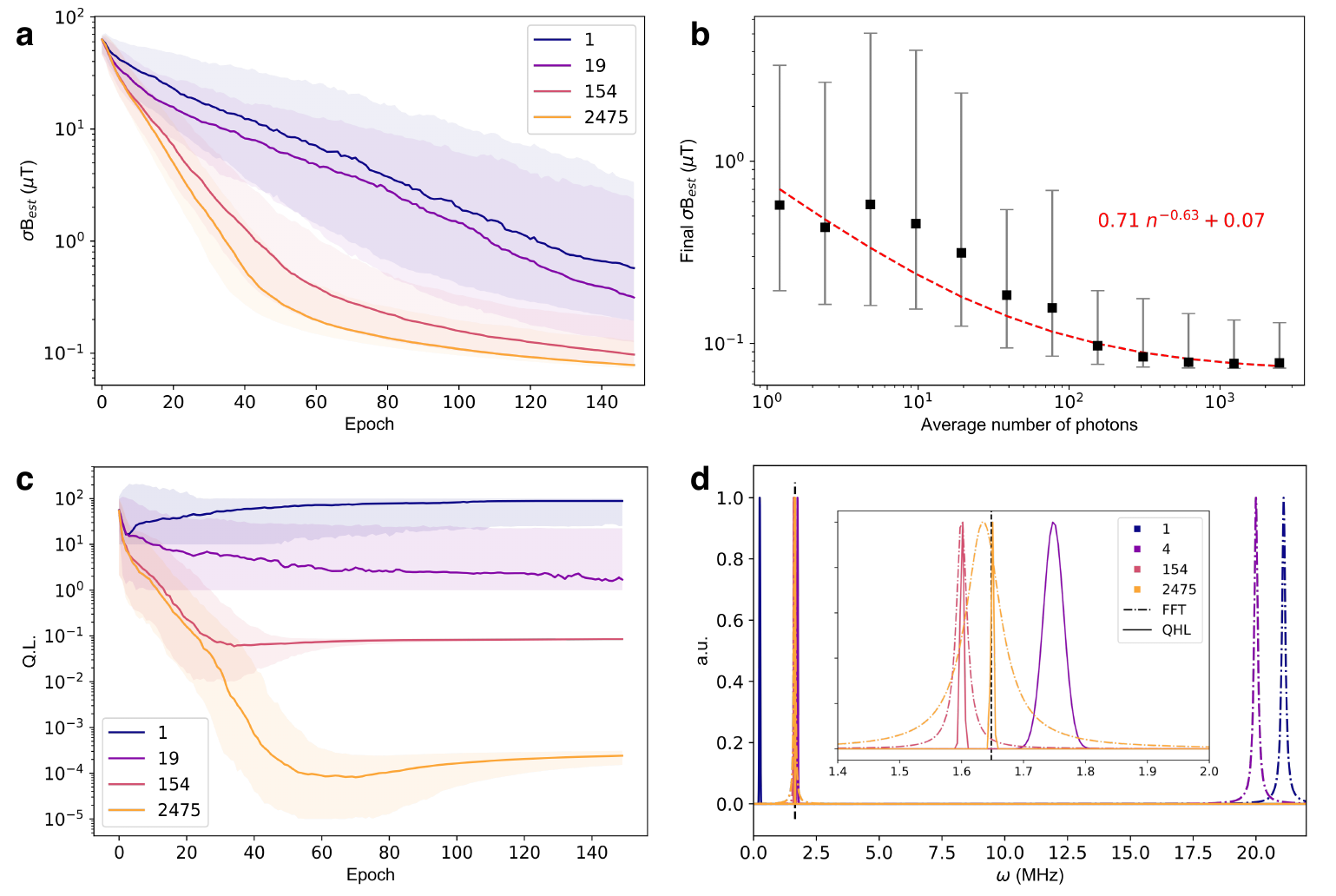} 
\caption{ 
\textbf{Key performances of MFL as the average number of photons collected $n_{\textrm{phot}}$ is increased}, estimated via 1000 independent runs of MFL, with an underlying lossless model (see Eq. 3 in Methods).
\textbf{a} Scaling of the median error estimate for a selection of cases,  as reported in the colour-coding legend.
\textbf{b} Final median error achieved by MFL for all the cases considered, along with a power-law fit (dashed line). Error bars indicate to 68.27 \% percentile ranges. 
\textbf{c} Scaling of the quadratic losses for the same representative runs as in a.
\textbf{d} Comparison of the final estimates for the Ramsey frequency provided for representative $ n_{\textrm{phot}}$ cases, by FFT and QHL methods, respectively in solid and dash-dotted lines.
In a\&c, shaded areas indicate the credible intervals corresponding to 68.27 \% percentile ranges.
}
\label{fig:SI_ScalingSweep}
\end{figure*}

\section*{Precision bounds and sensitivity}
\label{sec:sensitivity}

\newcommand{\data}{\texttt{data}}

In assessing the performance of MFL, it is helpful to compare the uncertainties achieved with those achieved by FFT, using the same datasets.
We begin by considering the Cram\'er--Rao bound~\cite{Cover:2006gz}.
Suppose that our procedure implements a function $\hat{B}(\data)$ of the entire data record ($\data$) that estimates the true magnetic field $B$.
After, we want to minimise the squared error $L = (\hat{B}(\data) - B)^2$ as much as possible.
If $\hat{B}$ as the average of $\hat{B}(\data) - B$ over all possible data records is zero, then we state that our procedure is unbiased.
In this case, the Cram\'er--Rao bound provides, that on average over all data records one finds~\cite{Ferrie:2014cf},
\begin{align}
    \label{eq:heisenberg-motivation}
    \mathbb{E}[L] & \ge \frac{1}{\sum_i \gamma^2 \tau_i^2},
\end{align}
where $\tau_i$ is the evolution time used at the $i$th step of the MFL procedure.
We stress that this inequality holds only on average; after all, we might be ``lucky'' with the estimate that we assign to any \emph{particular} data record.

The right-hand side of this inequality is derived using the Fisher information $I$ for a single measurement,
\begin{align}
    I(B, \tau)
        \mathrel{:=} {} &
            \mathbb{E}_{E} \left[
                \left(\partial_B \log P(E | \vec{x}, \tau)\right)^2
            \right] \\
        = {} &
            \tau^2 \gamma^2.
\end{align}
The Fisher information for an experiment consisting of multiple independent measurements is given by the sum of the Fisher informations for each measurement, giving the Cram\'er--Rao bound (Eq.~\ref{eq:heisenberg-motivation}).

Let $T$ be the total phase accumulation time used for a single ``run'' of a magnetometry procedure; in our case, $T = \sum_i \tau_i$.
By the above argument, $L$ can then scale no better than $T^{-2}$, corresponding to consolidating our phase accumulation into a single measurement.
This observation is sometimes referred to as the \emph{Heisenberg limit} for magnetometry.

At the other extreme, suppose that we have a total time budget of $T$, that we are able to spend on a magnetometry experiment, such that we can consider repeating a given procedure $N = T / \tau$ times.
The factor of $N$ then factors out of the Cram\'er--Rao bound, giving
\begin{align}
    \label{eq:sql-motivation}
    \mathbb{E}[L]
        & \ge \frac{1}{N \sum_i \gamma^2 \tau_i^2}
          \propto \frac{1}{T}.
\end{align}
The observation that $L \propto 1 / T$ is sometimes referred to as the \emph{standard quantum limit},in the case that we repeat a magnetometry procedure for $N$ independent iterations.
Indeed, we can use this observation to motivate a general figure of merit for the time budget of the a given magnetometry procedure.
Assume that the Fisher information for a given procedure is $I = N I_0$, where $I_0(\tau)$ is the Fisher information for a single repetition using phase accumulation time $\tau$.
Then it follows $I = T (I_0 / \tau)$.
Next, we define $\eta(T) \mathrel{:=} \sqrt{I_0(T) / \tau}$ as the \emph{sensitivity} of the proposed magnetometry procedure. 

Using this definition, we can then restate the standard quantum limit as the statement that $\eta^2(T)$ is \emph{constant} in $T$.
That is, a magnetometry procedure bound by the standard quantum limit gains no advantage from phase accumulation time beyond that conferred by repeating the entire procedure for $R$ independent runs.
By contrast, a Heisenberg limited magnetometry procedure has a sensitivity which scales as $\eta^2(T) \propto 1 / T$, indicating that an additional advantage can possibly be gained by using longer phase accumulation times.

So far we have considered the case in which $T^{*}_{2} \gg T$, such that we can approximate the dynamics of our magnetometry experiment as dephasing-free.
The dichotomy between the Heisenberg and standard quantum limit scalings, however, is changed by dephasing such that we have to consider the definition of the sensitivity in the dephasing-limited case.
In particular, Ref.~\cite{Ferrie:2014cf} derived that the Fisher information for $T^{*}_{2}$-limited magnetometry is given by
\begin{align}
    \label{eq:noisy-fisher}
    \mathbb{E}[L] & \ge
        \frac{
            \csc^2(B \gamma  t) \left(
                e^{2 \Gamma  t} -
                \left(e^{\Gamma  t}-1\right)^2 \cos ^2(B \gamma  t)
            \right)
        }{
            \gamma ^2 t^2 \left(e^{\Gamma  t}-1\right)^2
        },
\end{align}
where $\Gamma \mathrel{:=} 1 / T^{*}_{2}$ is used to represent dephasing in frequency units, in analogy with $\gamma B$.
We note, that unlike the Fisher information describing the noiseless case, the bound \autoref{eq:noisy-fisher} for the dephasing-limited case is \emph{not} independent of the true value of $B$.
Thus, to determine the achievable sensitivity in the case of dephasing-limited magnetometry, we must either assume a particular value of the field $B$ being estimated, or must generalise beyond the Cram\'er--Rao bound.
We choose the latter case in this work, which provides further insight into the trade-off between phase accumulation time and experimental repetitions for $\Gamma > 0$.

Specifically, we consider the van Trees inequality (also known as the Bayesian Cram\'er--Rao bound)~\cite{Gill:1995vn},
\begin{align}
    \mathbb{E}_{B,E}[L] \ge \frac{1}{J_\pi + \mathbb{E}_{B}[I(B)]},
\end{align}
where $J_\pi$ describes the error that can be achieved using prior information, and $\mathbb{E}_B$ an expectation value over a \emph{distribution} of different hypotheses about the field $B$.
We intentionally do not further define $J_\pi$, as this term depends on the context in which a magnetometry procedure is used, rather than on the magnetometry procedure itself.
Moreover, the effect of $J_\pi$ is minimal in the limit of large  experimental data sets, such that ineffectively consists of a correction to the Cram\'er--Rao bound in the case of finite data records \cite{Opper:1999uu}.

In analogy to the Fisher information derivation above, the field-averaged Fisher information in the dephasing-limited case gives $\mathbb{E}_B[I(B)] \le (\tau e^{-\tau \Gamma})^{2}$ for a single phase accumulation $\tau$.
Hence, the analogous bound to \autoref{eq:heisenberg-motivation} is given by
\begin{align}
    \label{eq:noisy-van-trees}
    \mathbb{E}_{B,E}[L] \ge \frac{1}{
        J_\pi + \sum_i (\gamma \tau e^{-\tau \Gamma})^2
    }.
\end{align}

To derive the sensitivity in the van Trees case, let $J_0(T) = \mathbb{E}_B[I_0(T, B)]$ be the average Fisher information for a dephasing-limited procedure.
We can then define the average sensitivity $\bar{\eta}(T) = J_0(T) / T$ for a total phase accumulation time $T$ to reformulate the van Trees inequality in a more practical form for our purposes,
thus
\begin{align}
    \mathbb{E}_{E,B}[L] \ge \frac{1}{J_\pi + T \bar{\eta}^2(T)}.
\end{align}
Following \autoref{eq:noisy-van-trees}, $\bar{\eta}^2(T)$ is constant if a fixed phase accumulation time is used, while $\bar{\eta}^2(T) \propto 1 / T$ if $T \Gamma \ll 1$.
The average Fisher information saturates at $T = T^{*}_{2}$, however, such that the Heisenberg and standard quantum limits coincide as $T$ approaches $T^{*}_{2}$.
Therefore, the performance 
observed in Fig.~\ref{Fig:Figure1}a is limited by saturation near $T = T^{*}_{2} \gtrapprox 0.1 \second$.

\section*{Absolute precision scaling}
\label{sec:SI_shots}

As discussed in the main text and Methods, using a room temperature set-up can be challenging for the effect of quantum projection noise 
and readout infidelities. These need to be properly addressed when reduced sequence repetitions lead to a low number of PL photons to be detected when recording a fringe. 
The results in terms of absolute scaling $\bar \eta$ have already been discussed (see e.g. Fig.~3 in the main paper). Here, we complete those analyses with additional studies. 
In Fig.~\ref{fig:SI_ScalingSweep}a\&b, we report respectively the scaling and ultimate uncertainty achievable by MFL after 150 epochs for a subset of cases with $n_{\textrm{phot}} = 1, ... \; 2475$. Fig.~\ref{fig:SI_ScalingSweep}b suggests an approximate $\propto 1/n_{\textrm{phot}}$ gain  in the uncertainty achievable halting the protocol after a fixed number of steps. 
From Fig.~\ref{fig:SI_ScalingSweep}c we observe that the choice for $N$ in this case is motivated by running MFL for enough steps, to observe for all cases the convergence of the median quadratic losses -- i.e. the square error in the parameters' estimate, here $Q.L. := (B-B_{\textrm{est}})^2$. To estimate the true $B$, we run MFL once over the whole dataset ($M=M^{tot}=20275$), checking that the result is consistent with FFT.\\
We remark how the advantages in increasing the $n_{\textrm{phot}}$ used (along with the higher precision scaling that increases from $\bar{T}^{-0.73}$ for $n_{\textrm{phot}}=4$ to Heisenberg limited for $n_{\textrm{phot}} \geq 20$) come at the expense of worse final absolute sensitivities achievable by the protocol. This is due to the linear increment of experimental overheads with $n_{\textrm{phot}}$. \\
The robustness against sources of noise present in the room temperature set-up is emphasized by Fig.~\ref{fig:SI_ScalingSweep}d, where we plot the estimates obtained by MFL for a similar subset of $n_{\textrm{phot}}$'s analysed. We observe that for $n_{\textrm{phot}} \geq 4$, MFL estimates are all substantially consistent with the result obtained for $M=M^{tot}$, within the estimated uncertainty $\sigma(B_{\textrm{est}})$ and taking into account minor fluctuations in $B$ that might have occurred during the collection of the whole dataset. By contrast, we observe how FFT estimates are completely unreliable at the noise level corresponding to $\mathcal{O} (n_{\textrm{phot}}) = 10 $.

\begin{figure*}[ht!]
\centering
\includegraphics[width=0.95\textwidth]{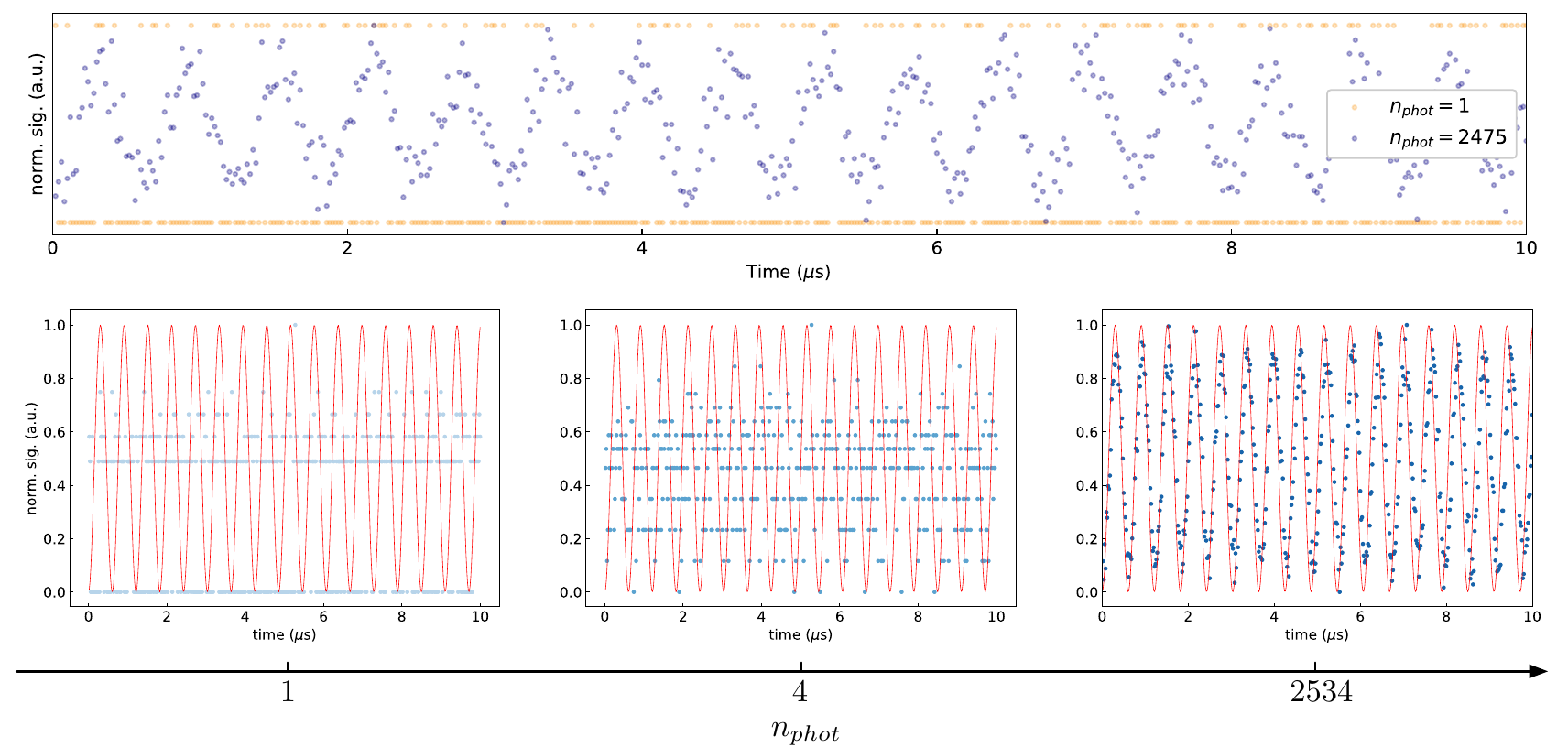} 
\caption{\textbf{Ramsey sets mediated through different numbers of single sequences}, corresponding to the various $n_{\textrm{phot}}$ reported on the axis below. Experimental data (as blue dots) are reported together with a sinusoidal fit obtained from the case $n_{\textrm{phot}} = 2534$ (as red lines).
The unbalance towards the $0$ measurement outcome is evident in the cases $n_{\textrm{phot}}=1, 4$. Data-points whose normalisation is higher than $0.5$ correspond to Poissonian distributed multi-photon events still present in this case.   
}
\label{fig:SI_FringeSweeps}
\end{figure*}

\section*{The role of noise for low PL photon counts}
\label{sec:SI_noise}

Finally, we observe how for $n_{\textrm{phot}} < 4$, the Bayesian process fails due to increased experimental noise and reduced statistics, underestimating both the real $\omega$ and the uncertainty associated with it. For example, this is evident from  Fig.~\ref{fig:SI_ScalingSweep}c, as the $Q.L.$ does not improve with the number of epochs.
In particular, losses in the system cause an asymmetry between $\xi_0$ and $\xi_1$, respectively the overall readout fidelities for the states $\ket{0}$ and $\ket{1}$ (i.e. taken all sources of noise and loss into account). From experimental raw data for $n_{\textrm{phot}} = 1$ (see Fig.~\ref{fig:SI_FringeSweeps}), we observed that if we assume $\xi_0 \sim 1$, then $\xi_1 \simeq 0.54$. This translates in unbalanced output probabilities, that conflict with the underlying assumption made so far of a binomial model for the outcomes $E \in \{0,1\}$, with probabilities given by the likelihood in Eq.~\ref{eq:likeT2}. This level of ``poisoning'' in the assumed model is evidently beyond the CLE noise robustness~\cite{Granade:2017kb}.  

In order to prove there is no fundamental limit preventing MFL to provide correct estimates, within uncertainty, given a correct model, we thus modified the likelihood such that:
\begin{equation}
 \mathcal{L}'(1|\vec x, \tau) = \xi \mathcal{L}(1|\vec x, \tau)
 \label{eq:SI_likeLOSS}
\end{equation}
where $\xi \in [0,1]$, $\mathcal{L}'(0|\vec x, \tau) = 1 - \mathcal{L}'(1|\vec x, \tau)$ and for $\xi = 1$ we recover the usual $\mathcal{L}'(0|\vec x, \tau) = \mathcal{L} (0| \vec x; \tau)$ of Eq.~\ref{eq:likeT2}. In order to estimate $\xi$, we use it as the free parameter in a preliminary CLE run against the same dataset, but assuming $\omega$ known from the inference process with $M = 20275$ (i.e. having $\vec x = \{ \xi \}$). We thus obtain $\xi = 0.72$, and use this as a known parameter when running MFL with $\mathcal{L}'(B)$. In principle, $\xi$ could also be estimated from a multi-parameter inference model. 

\begin{figure*}[ht!]
\centering
\includegraphics[width=0.85\textwidth]{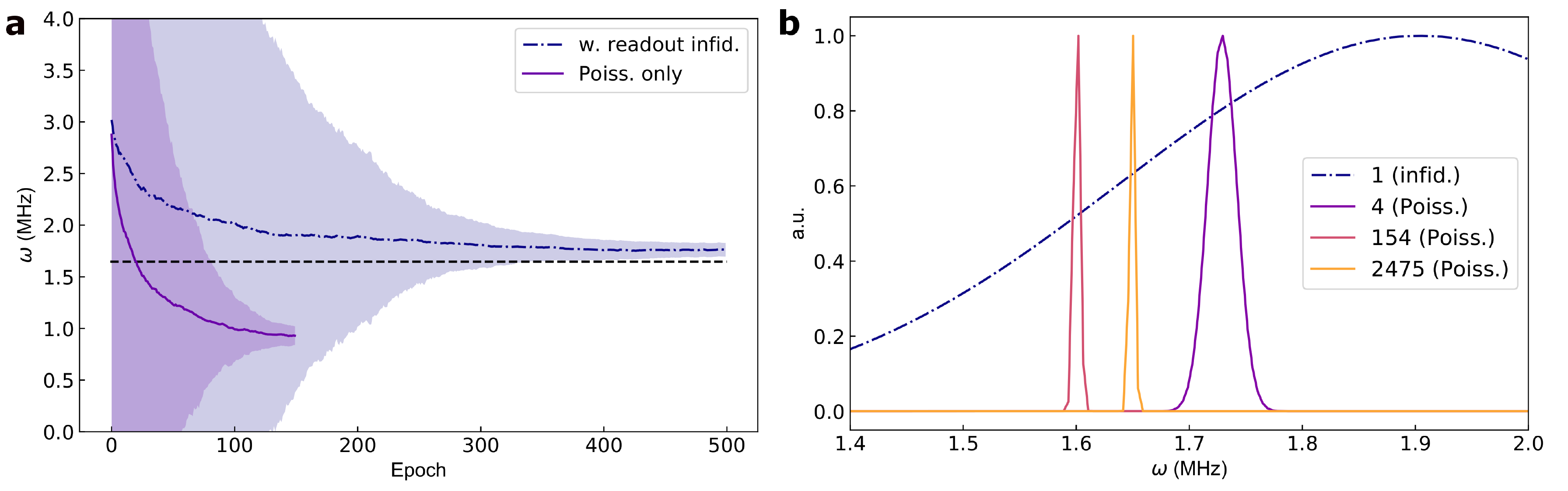} 
\caption{
\textbf{Noise-compensation in the inference process.}
\textbf{a}, Estimate of the precession frequency $\omega$ from the set $\epsilon_1$, using $M = 8 \Rightarrow n_{\textrm{phot}} =1$ average photons collected per step, in peak configuration. In violet the result adopting the usual likelihood in Eq.~\ref{eq:likeT2}, capable of handling only Poissonian noise in the the data. In blue, results from the modified likelihood Eq.~\ref{eq:SI_likeLOSS}, allowing for an extended number of epochs. Shaded areas here represent the median $\sim 68.27$ \% credible interval provided intrinsically by MFL at each epoch, averaged over 1000 runs.
\textbf{b}, Estimates for $\omega$, and uncertainties as a Gaussian fit over the learnt posterior, for the two cases in \textbf{a}, along with some other representative runs from Fig.~\ref{fig:SI_ScalingSweep}, after 150 epochs. The inference process with no model for infidelities in place, and  $n_{\textrm{phot}} =1$, falls outside of the plotted interval. 
}
\label{fig:SI_noise}
\end{figure*}

The result is reported in Fig.~\ref{fig:SI_noise}. Intuitively, measurement outcomes are interpreted as less informative by the inference process, as $E=0$ might be due to additional losses. This effectively slows down the learning rate per-epoch, but at the same time restores a correct behaviour of MFL   .

\section*{Wide range operability of MFL methods}
\label{sec:SI_range}

It is known how in Ramsey experiments, adaptive choices of time can lead not only to scalings beyond the standard quantum limit, but also to improved dynamic ranges for the sensed magnetic field $B$, up to $ \omega_{max} (B) / \sigma(\omega) \leq \pi T / \tau_{min}$. 
Given that MFL is adaptive in the choices of the phase accumulation time $\tau$, and we have shown that its precision scaling is Heisenberg limited, it follows naturally that also MFL benefits from the high-dynamic range already reported by previous experiments. 

In the main paper, we already showed applicability of MFL for cases in the dataset $\epsilon_3$, with $B \in [8.3, 550]$ \unit{}{\micro \tesla} (see Fig.~2 in the main text).
Here we complement this study with an additional case ($\alpha_2$) exhibiting $B_h=$\unit{713}{\micro \tesla}. In the case of this dataset, equivalently to $\alpha_1$, $M \sim 20000$ single sequences were collected and averaged from the experimental set-up, so it was reasonable to adopt a majority voting scheme to use the additional information in the data. We stress that such high intensities of $B$ tend to make least-squares fit procedures with no initial guess of the parameters fail.

\begin{figure*}[ht!]
\centering
\includegraphics[width=0.85\textwidth]{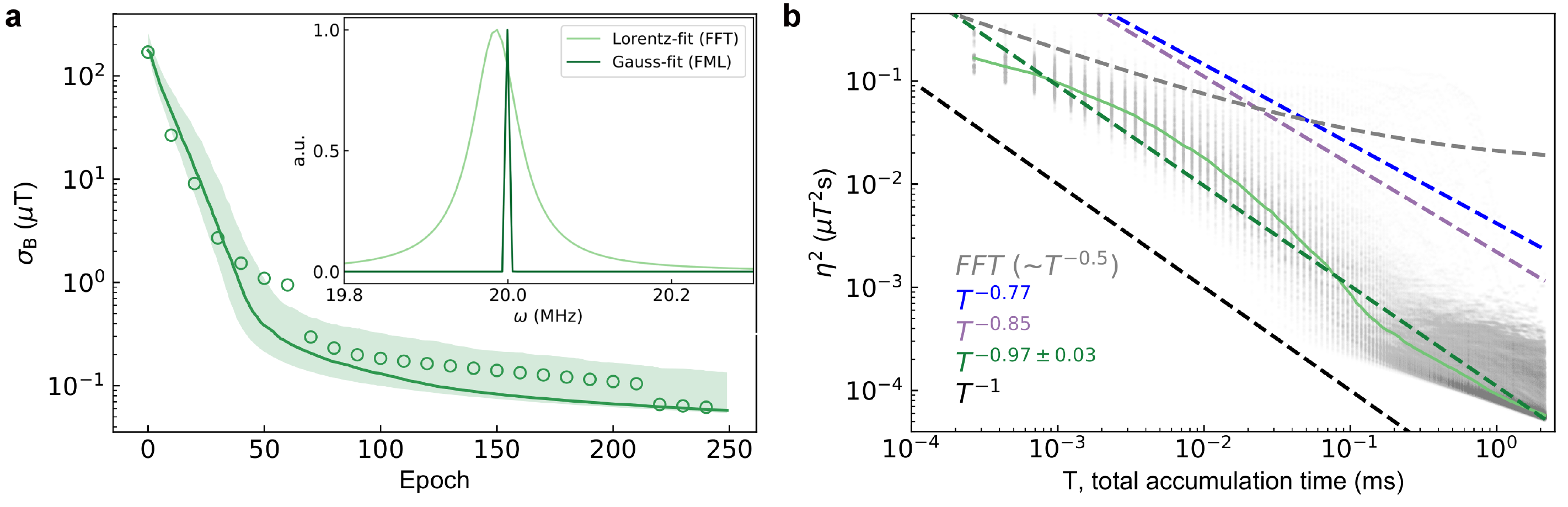} 
\caption{
\textbf{Sensing high-intensity magnetic fields} ($B \simeq$ \unit{710}{\micro \tesla}) with $\tau_{max} \ll T^{*}_{2}$. 
\textbf{a}, Scaling in the median uncertainty over 1000 protocol runs performed each on a random Ramsey set, among those available in the ensemble $\alpha_2$ of Supplementary Table 1.
\textbf{b}, Precision scaling for MFL (in green), calculated over the same ensemble, compared with previous approaches (purple, blue) and the Heisenberg limited scaling (black). The error associated to the scaling is estimated via a bootstrap technique. The results from single runs are reported as a density plot in green. All offsets for clarity. (See for comparison and further details Fig.~2a\&b in the main text.)
}
\label{fig:SI_HighField}
\end{figure*}

The results in terms of $\sigma(B_{\textrm{est}})$ and precision $\eta^2$ scalings are reported in Fig.~\ref{fig:SI_HighField}. We observe how after 250 epochs, the difference in the final uncertainties provided by MFL is $|\sigma (B_{\alpha_1}) - \sigma (B_{\alpha_2})| \simeq 10^{-3}$ {\micro \tesla}. It can thus be considered approximately independent of the strength of the magnetic field. Also the precision scaling is the same, within error, of the one observed for the lower field in $\alpha_2$.

\section*{Tracking}
\label{sec:SI_tracking}

\begin{figure*}[b!]
\centering
\includegraphics[width=0.95\textwidth]{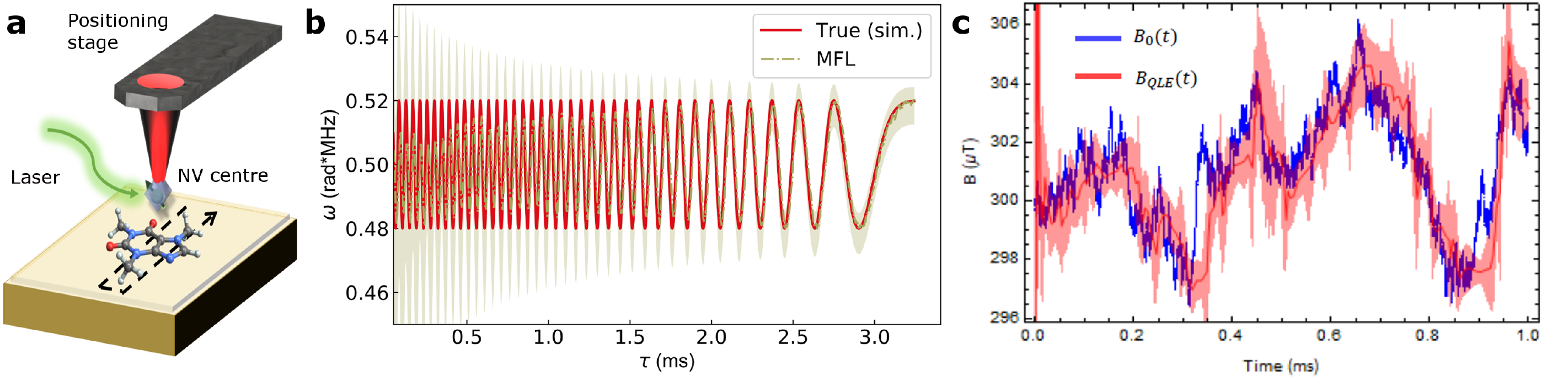} 
\caption{\textbf{Magnetic field on-line tracking via MFL.}  
\textbf{a}, Pictorial representation of possible applications of a tracking protocol, where an NV centre positioned at the end of a scanning microscope is used to scan the magnetic field $B$ in the proximity of a molecule absorbed on a substrate. 
\textbf{b}, Simulation of MFL capabilities to track a chirped sinusoidal signal, with no experimental overhead and only Poissonian noise present in the data (i.e. high-fidelity readout). The frequency $\omega$ is linearly increased after each update step. 
\textbf{c}, Average performance, mediated over 1000 independent runs, of CLE tracking a magnetic field undergoing an Ornstein-Uhlenbeck stochastic process. In \textbf{b\&c}, shaded areas corresponds to the usual $\sim 68$ \% percentile credible range adopted in this paper. 
}
\label{fig:SI_MagneticFieldTracking}
\end{figure*}

In the main text, we tested against experimental data the tracking capabilities of the MFL protocol. In Fig.~4 we reported the results in the case where the magnetic field intensity $B$ is synthetically altered stepwise, at random times, in a fashion completely equivalent to a stochastic time-dependent Poisson process $\mathcal{P} (t)$. 
Such random, abrupt variations in the magnetic field might for example reproduce applicative scenarios such as the raster scanning of a surface embedding magnetic nanoparticles. A sketch of a possible experimental set-up is provided in Fig.~\ref{fig:SI_MagneticFieldTracking}a. The modifications to the standard CLE inference process required by this particularly demanding tracking scenario are summarised as pseudocode in Algorithm~\ref{algo:SI_trackcode}. The modifications to the standard inference process adopted amount to detect changes in the sensed parameter, that completely invalidate the current posterior, and thus suggest a reset of the prior as the most effective update step. Without triggering such reset events, huge stepwise changes would otherwise require a long time for MFL to react, because of the little support provided by the prior to the new value. 

In Fig.~\ref{fig:SI_MagneticFieldTracking}b, we show a simulated performance of MFL in a representative run with time-varying $B$. The figure exemplifies the decrease in the rate of failure events, as the frequency of the oscillating signal is decreased with time. We loosely define failure events, all those $\tau_i$ at which the quadratic loss of a single run $Q.L. (\tau_i) \gg \sum_r^{R} [ Q.L. (\tau_i, r) ] / R$, the mean performance achievable by the protocol, estimated across $R \gg 1$ independent runs. We modify synthetically the magnetic field in the simulations as $B(\tau) = B_0 + b \cos (\nu \tau) $ with $b \ll B_0$, equivalently to Fig.~4c of the main text, but in this case we chirp the oscillating frequency for each run, and thus $\nu (\tau) = \nu_0 - k \tau$, with $\nu_0$ and $k$ constants. 
We notice how points where the second derivative of the oscillating magnetic signal is highest are those where failure events tend to occur.

Finally, in Fig.~\ref{fig:SI_MagneticFieldTracking}c, we display the performance expected for MFL when tracking a brownian-like varying signal. Here the `true' signal is numerically simulated according to an Ornstein-Uhlenbeck process, similarly to the theoretical analysis in \cite{Bonato:2017fr}.

\begin{algorithm}
\hrulefill
\vskip-8pt
\caption{MFL algorithm with stepwise change detection}
\vskip-7pt
\hrulefill
\label{mflmod}
\begin{algorithmic}
\smallskip

\STATE  \textbf{Input} :
An initial prior distribution $\pi_{ini}$ over models.
\STATE  \textbf{Input} (additional) : $\tt{r}_{resample}$ \indent $\triangleright$ a rate parameter estimating how many $ epochs $ occur before \textsc{Resample} is called
\STATE  \textbf{Input} (additional) : $\tt{p}_{reset}$ \indent $ \triangleright $ parameter adjusting the frequency of posterior-reset events

\STATE \indent 
  \textbf{function} \textsc{EstimateAdaptive} (n, $\pi_{ini}$, N, $\tt{a}$ (the resampling parameter), 				$\tt{t}_{resample}$ (the $\tt{resample\_threshold}$), 
  \textsc{Optimize}, \textsc{Util}, n\textsubscript{guesses},
  \textsc{GuessExperiment}):
    \STATE \indent \indent
    	$w_i \leftarrow 1/n$
    \STATE \indent \indent
    	draw each $\omega_i$ independently from $\pi_{ini}$
    \STATE \indent \indent    
        $l_{evres} \leftarrow 0$
    \STATE \indent \indent    
        $l_{evstp} \leftarrow 0$
    \STATE \indent \indent    
        initialize $\vec x = [\bar x_1, ... , \bar x_N ]$
    \STATE \indent \indent
    \textbf{for} $i_{exp} \in 1 ... N$ \textbf{do}:
    \STATE \indent \indent \indent
    	... \\

\STATE \indent \indent \indent 
    	\textbf{if} $\sum_i w_i^2 < N \tt{t}_{resample}$:
        \indent $\triangleright$ if the effective sample size is below the threshold
        
    \STATE \indent \indent \indent \indent 	
    	\textbf{if} $|i_{exp}- l_{evres}| \geq \tt{r}_{resample}$ \textbf{OR} $|i_{exp}- l_{evstp}| \leq \tt{p_{reset}}$:
        \indent $\triangleright$ resample as usual
        
    \STATE \indent \indent \indent \indent \indent    
        $\{w_i\}, \{x_i\} \leftarrow$ \textsc{Resample}(
        	$\{w_i\}, \{x_i\}, \tt{a}$ )
    \STATE \indent \indent \indent \indent \indent 
        $l_{evres} \leftarrow i_{exp}$
     	\indent $\triangleright$ store last resampling event
    \STATE \indent \indent \indent \indent
    \textbf{else}:  
    \indent $\triangleright$ reset the procedure
    
    \STATE \indent \indent \indent \indent \indent
    	$\pi \leftarrow \pi_{ini}$
    \STATE \indent \indent \indent \indent \indent
    	$w_i \leftarrow 1/n$
    \STATE \indent \indent \indent \indent \indent
    	draw each $\omega_i$ independently from $\pi_{ini}$  
    \STATE \indent \indent \indent \indent \indent
    	 $l_{evstp} \leftarrow i_{exp}$
         \indent $\triangleright$ store last reset event
    \STATE \indent \indent \indent \indent \indent 	
       	 \textbf{continue} from $i_{exp}+1$
    \STATE \indent \indent \indent \indent 
    	\textbf{end if}
    \STATE \indent \indent \indent
    	\textbf{end if}
        
    \STATE \indent \indent \indent
        ...
    \STATE \indent \indent \indent
    $\bar x_i \leftarrow$ \textsc{Mean} $(\{w_i\}, \{x_i\})$
    \indent $\triangleright$ append the new estimate from the mean
    \STATE \indent \indent
    	\textbf{end for}    
    
    \STATE \indent  \textbf{end function} \\     

    \STATE  \textbf{Output:} $\vec x$, storing the instantaneous values of the unknown parameter

\end{algorithmic}
\label{algo:SI_trackcode}
\end{algorithm}

\end{document}